\documentclass[
  journal=pasa,
  manuscript=research-article,
  year=2025,
  volume=37,
]{cup-journal}

\usepackage{amsmath}
\usepackage{amssymb}
\usepackage[nopatch]{microtype}
\usepackage{booktabs}
\usepackage[acronym]{glossaries-extra}
\usepackage{natbib}
\usepackage{upgreek}
\usepackage{xspace}
\usepackage[normalem]{ulem}
\usepackage{caption}
\usepackage{subcaption}
\usepackage{nicefrac}
\usepackage{xfrac}
\usepackage{rotating}
\usepackage{tabularray}
\usepackage{orcidlink}
\usepackage[textsize=footnotesize]{todonotes}
\setuptodonotes{inline}

\usepackage{tikz}
\usetikzlibrary{shapes.geometric, arrows}
\tikzstyle{startstop} = [rectangle, rounded corners, minimum width=2cm, minimum height=1cm, text centered, draw=black]
\tikzstyle{process} = [rectangle, minimum width=2cm, minimum height=1cm, text centered, draw=black]
\tikzstyle{decision} = [diamond, minimum width=2cm, minimum height=1cm, text centered, draw=black]
\tikzstyle{arrow} = [thick,->,>=stealth]

\usepackage{listings}
\definecolor{codegreen}{rgb}{0,0.6,0}
\definecolor{codegray}{rgb}{0.5,0.5,0.5}
\definecolor{codepurple}{HTML}{C42043}
\definecolor{backcolour}{HTML}{F2F2F2}
\definecolor{bookColor}{cmyk}{0,0,0,0.90}
\lstdefinestyle{sqlstyle}{
        backgroundcolor=\color{backcolour},   
        commentstyle=\color{codegreen},
        keywordstyle=\color{codepurple},
        numberstyle=\footnotesize\color{codegray},
        stringstyle=\color{codepurple},
        basicstyle=\footnotesize\ttfamily,
        breakatwhitespace=false,
        breaklines=true,
        captionpos=b,
        keepspaces=true,
        numbers=left,
        numbersep=5pt,
        showspaces=false,
        showstringspaces=false,
        showtabs=false
}
\lstdefinestyle{pythonstyle}{
    keywordstyle=\color{DarkOrchid3},
    commentstyle={\color{Green4}\itshape\sffamily},
    numberstyle=\tiny\color{Ivory4},
    stringstyle=\color{Purple1},
    basicstyle=\sffamily\footnotesize,
    breakatwhitespace=false,
    breaklines=true,
    captionpos=b,
    keepspaces=true,
    numbers=left,
    numbersep=5pt,
    showspaces=false,
    showstringspaces=false,
    showtabs=false,
    tabsize=2
}

\graphicspath{ {./images/} }

\usepackage{hyperref}
\hypersetup{
    colorlinks=True,
    pdftitle={A Census of Double-Peaked Lyman-alpha Emitters in MAGPI:
Classification, Global Characteristics, and Spatially Resolved Properties - Tamal Mukherjee},
    pdfauthor={Tamal Mukherjee, Macquarie University},
    linkcolor=black,
    urlcolor=blue,
    citecolor=blue
    }

\setabbreviationstyle[acronym]{long-short}
\glsdisablehyper

\newacronym{nir}{near-IR}{near-infrared}
\newacronym{sso}{SSO}{Siding Springs Observatory}
\newacronym{rsaa}{RSAA}{Research School of Astronomy and Astrophysics}
\newacronym{sami}{SAMI}{Sydney-AAO Multi-object Integral field spectrograph}
\newacronym{aat}{AAT}{Anglo-Australian Telescope}
\newacronym{ifs}{IFS}{integral-field spectroscopy}
\newacronym{uv}{UV}{ultraviolet}
\newacronym{sed}{SED}{spectral energy distribution}
\newacronym{ir}{IR}{infrared}
\newacronym{sfr}{SFR}{star formation rate}
\newacronym{mw}{MW}{Milky Way}
\newacronym{ifu}{IFU}{integral field unit}
\newacronym{gama}{GAMA}{Galaxy and Mass Assembly}
\newacronym{dr3}{DR3}{third data release}
\newacronym{snr}{SNR}{signal-to-noise ratio}
\newacronym{agn}{AGN}{active galactic nuclei}
\newacronym{dig}{DIG}{diffuse ionised gas}
\newacronym{magpi}{MAGPI}{Middle Ages Galaxy Properties with Integral field spectroscopy}
\newacronym{manga}{MaNGA}{Mapping Nearby Galaxies at Apache Point Observatory}
\newacronym{3dhst}{3D-HST}{Three-Dimensional Hubble Space Telescope}
\newacronym{califa}{CALIFA}{Calar Alto Legacy Integral Field Area}
\newacronym{muse}{MUSE}{Multi Unit Spectroscopic Explorer}
\newacronym{hudf}{HUDF}{Hubble Ultra-Deep Field}
\newacronym{glao}{GLAO}{ground-layer adaptive optics}
\newacronym{lgs}{LGS}{laser guide star}
\newacronym{pmas}{PMAS}{Potsdam Multi Aperture Spectrograph}
\newacronym{jades}{JADES}{JWST Advanced Deep Extragalactic Survey}
\newacronym{ism}{ISM}{interstellar medium}
\newacronym{cirs}{CIRS}{Cluster Infall Regions in the SDSS}
\newacronym{2dfgrs}{2dFGRS}{2dF Galaxy Redshift Survey}
\newacronym{ssfr}{sSFR}{specific star formation rate}
\newacronym{dar}{DAR}{differential atmospheric refraction}
\newacronym{sps}{SPS}{stellar population synthesis}
\newacronym{liner}{LINER}{low-ionisation nuclear emission-line region}
\newacronym{lae}{LAE}{Lyman-$\upalpha$ emitter}
\newacronym{jwst}{JWST}{James Webb Space Telescope}
\newacronym{gist}{GIST}{Galaxy IFU Spectroscopy Tool}

\def\ha{H$\alpha$}

\def\lya{Ly$\alpha$}

\def\oii{[O~{\sc ii}]}

\def\hi{H~{\sc i}}
\def\hii{H~{\sc ii}}

\def\ciii{C~{\sc iii}]}

\def\bt{$F_{\mathrm{blue}}\, / F_{\mathrm{total}}$}
\def\fesc{$f^{\mathrm{LyC}}_{\mathrm{esc}}$}

\title{A Census of Double-Peaked \lya\ Emitters in MAGPI:
Classification, Global Characteristics, and Spatially Resolved Properties}

\author{Tamal Mukherjee
\textsuperscript{\orcidlink{0009-0004-7639-869X}}}
\affiliation{School of Mathematical and Physical Sciences, Macquarie University, NSW 2109, Australia}
\alsoaffiliation{ARC Centre of Excellence for All Sky Astrophysics in 3 Dimensions (ASTRO 3D)}
\alsoaffiliation{Astrophysics and Space Technologies Research Centre, Macquarie University, Sydney, NSW 2109, Australia}

\email[T. Mukherjee]{tamal.mukherjee@hdr.mq.edu.au}

\author{Tayyaba Zafar
\textsuperscript{\orcidlink{0000-0003-3935-7018}}}
\affiliation{School of Mathematical and Physical Sciences, Macquarie University, NSW 2109, Australia}
\alsoaffiliation{ARC Centre of Excellence for All Sky Astrophysics in 3 Dimensions (ASTRO 3D)}
\alsoaffiliation{Astrophysics and Space Technologies Research Centre, Macquarie University, Sydney, NSW 2109, Australia}

\author{Themiya Nanayakkara
\textsuperscript{\orcidlink{0000-0003-2804-0648}}}
\affiliation{Centre for Astrophysics and Supercomputing, Swinburne University of Technology, PO Box 218, Hawthorn 3122, VIC, Australia}
\alsoaffiliation{ARC Centre of Excellence for All Sky Astrophysics in 3 Dimensions (ASTRO 3D)}

\author{Siddhartha Gurung-Lopez
\textsuperscript{\orcidlink{0000-0001-9333-8470}}}
\affiliation{Observatori Astron\`omic de la Universitat de Val\`encia, Ed. Instituts d’Investigaci\'o, Parc Cient\'ific. C/ Catedr\'atico Jos\'e Beltr\'an, n2, 46980 Paterna, Valencia, Spain
}
\alsoaffiliation{Departament d’Astronomia i Astrof\'isica, Universitat de Val\`encia, 46100 Burjassot, Spain}

\author{Anshu Gupta
\textsuperscript{\orcidlink{0000-0002-8984-3666}}}
\affiliation{International Centre for Radio Astronomy Research (ICRAR), Curtin University, Bentley, WA, Australia
}
\alsoaffiliation{ARC Centre of Excellence for All Sky Astrophysics in 3 Dimensions (ASTRO 3D)}

\author{Scott M. Croom
\textsuperscript{\orcidlink{0000-0003-2880-9197}}}
\affiliation{Sydney Institute for Astronomy, School of Physics, University of Sydney, NSW 2006, Australia
}
\alsoaffiliation{ARC Centre of Excellence for All Sky Astrophysics in 3 Dimensions (ASTRO 3D)}

\author{Andrew Battisti
\textsuperscript{\orcidlink{0000-0003-4569-2285}}}
\affiliation{International Centre for Radio Astronomy Research (ICRAR), The University of Western Australia, Crawley, WA 6009, Australia}
\alsoaffiliation{Research School of Astronomy and Astrophysics, Australian National University, Canberra, ACT 2611, Australia}
\alsoaffiliation{ARC Centre of Excellence for All Sky Astrophysics in 3 Dimensions (ASTRO 3D)}

\author{Karl Glazebrook}
\affiliation{Centre for Astrophysics and Supercomputing, Swinburne University of Technology, PO Box 218, Hawthorn 3122, VIC, Australia}
\alsoaffiliation{JWST Australian Data Centre (JADC), Swinburne Advanced Manufacturing and Design Centre (AMDC), John Street, Hawthorn, VIC 3122, Australia}

\author{Polychronis Papaderos
\textsuperscript{\orcidlink{0000-0002-3733-8174}}}
\affiliation{Centro de Astrof´ısica e Ciencias ˆ do Espac¸, Universidade de Lisboa - OAL, Tapada da Ajuda, PT1349-018 Lisboa, Portugal}
\alsoaffiliation{Instituto de Astrof´ısica e Ciencias ˆ do Espac¸o - Centro de Astrof´ısica da Universidade do Porto, Rua das Estrelas, 4150-762 Porto, Portugal}

\author{Melissa Riggs}
\affiliation{School of Mathematical and Physical Sciences, Macquarie University, NSW 2109, Australia}
\alsoaffiliation{Astrophysics and Space Technologies Research Centre, Macquarie University, Sydney, NSW 2109, Australia}

\author{Emily Wisnioski
\textsuperscript{\orcidlink{0000-0003-1657-7878}}}
\affiliation{Research School of Astronomy and Astrophysics, Australian National University, Canberra, ACT 2611, Australia}
\alsoaffiliation{ARC Centre of Excellence for All Sky Astrophysics in 3 Dimensions (ASTRO 3D)}

\author{Caroline Foster
\textsuperscript{\orcidlink{0000-0003-0247-1204}}}
\affiliation{School of Physics, University of New South Wales, Sydney, NSW 2052, Australia}
\alsoaffiliation{ARC Centre of Excellence for All Sky Astrophysics in 3 Dimensions (ASTRO 3D)}

\author{Katherine E. Harborne
\textsuperscript{\orcidlink{0000-0002-2043-7985}}}
\affiliation{Institute for Computational Cosmology, Durham University, South Road, Durham DH1 3LE, UK
}
\alsoaffiliation{Centre for Extragalactic Astronomy, Durham University, South Road, Durham DH1 3LE, UK}
\alsoaffiliation{Department of Physics, Durham University, South Road, Durham DH1 3LE, UK}

\author{Claudia D. P. Lagos
\textsuperscript{\orcidlink{0000-0003-3021-8564}}}
\affiliation{International Centre for Radio Astronomy Research (ICRAR), The University of Western Australia, Crawley, WA 6009, Australia}
\alsoaffiliation{ARC Centre of Excellence for All Sky Astrophysics in 3 Dimensions (ASTRO 3D)}

\author{J. Trevor Mendel
\textsuperscript{\orcidlink{0000-0002-6327-9147}}}
\affiliation{Research School of Astronomy and Astrophysics, Australian National University, Canberra, ACT 2611, Australia}
\alsoaffiliation{ARC Centre of Excellence for All Sky Astrophysics in 3 Dimensions (ASTRO 3D)}

\author{Jahang Prathap
\textsuperscript{\orcidlink{0009-0004-0251-2672}}}
\affiliation{School of Mathematical and Physical Sciences, Macquarie University, NSW 2109, Australia}
\alsoaffiliation{ARC Centre of Excellence for All Sky Astrophysics in 3 Dimensions (ASTRO 3D)}
\alsoaffiliation{Astrophysics and Space Technologies Research Centre, Macquarie University, Sydney, NSW 2109, Australia}

\author{Stefania Barsanti
\textsuperscript{\orcidlink{0000-0002-9332-5386}}}
\affiliation{Sydney Institute for Astronomy, School of Physics, University of Sydney, NSW 2006, Australia}
\alsoaffiliation{Research School of Astronomy and Astrophysics, Australian National University, Canberra, ACT 2611, Australia}
\alsoaffiliation{ARC Centre of Excellence for All Sky Astrophysics in 3 Dimensions (ASTRO 3D)}

\author{Sarah M. Sweet
\textsuperscript{\orcidlink{0000-0002-1576-2505}}}
\affiliation{School of Mathematics and Physics, University of Queensland, Brisbane, QLD 4072, Australia}
\alsoaffiliation{ARC Centre of Excellence for All Sky Astrophysics in 3 Dimensions (ASTRO 3D)}

\author{Lucas M. Valenzuela
\textsuperscript{\orcidlink{0000-0002-7972-9675}}}
\affiliation{Universitäts-Sternwarte, Fakultät für Physik, 
Ludwig-Maximilians-Universität München, Scheinerstr. 1, 81679 München, Germany}

\author{Anilkumar Mailvaganam
\textsuperscript{\orcidlink{0009-0003-1221-1630}}}
\affiliation{School of Mathematical and Physical Sciences, Macquarie University, NSW 2109, Australia}
\alsoaffiliation{ARC Centre of Excellence for All Sky Astrophysics in 3 Dimensions (ASTRO 3D)}
\alsoaffiliation{Astrophysics and Space Technologies Research Centre, Macquarie University, Sydney, NSW 2109, Australia}


\begin{document}

\begin{abstract}
Double-peaked Lyman-$\alpha$ (\lya) profiles provide critical insights into gas kinematics and the distribution of neutral hydrogen (\hi) from the interstellar to the intergalactic medium (ISM to IGM), and serve as valuable diagnostics of ionising Lyman continuum (LyC) photon escape. We present a study of the global and spatially resolved properties of double-peaked \lya\ emitters (LAEs), based on VLT/MUSE data from the Middle Ages Galaxy Properties with Integral Field Spectroscopy (MAGPI) survey. From a parent sample of 417 LAEs at $z = 2.9 - 6.6$ in the first 35 fields, we identify 108 double-peaked LAEs using an automated peak classification technique. We measure a double-peak fraction of $\sim37\%$ at $z < 4$, decreasing to $\sim14\%$ at $z > 4$, likely due to enhanced IGM attenuation.
Approximately 17\% of the double-peaked LAEs are blue-dominated, possibly tracing gas inflows, though backscattering remains a viable alternative for sources without systemic redshift. The blue-to-total flux ratio exhibits a luminosity dependence: fainter lines generally show higher blue flux, though a few luminous sources also show strong blue peaks. We find a significant narrowing of the red peak at $z>4$, despite the presence of the blue peak, indicating intrinsic galaxy evolution rather than an effect of IGM attenuation.
Several LAEs exhibit residual flux in the absorption trough, with normalised trough flux anticorrelating with peak separation, reflecting variations in \hi\ column density.  We further investigate spatially-resolved properties of ten red-dominated LAEs with extended \lya\ halos. Despite azimuthal variations, both the blue-to-total flux ratio and normalised trough flux density increase with radius, while peak separation decreases. The red peak asymmetry shows only minor radial changes. These trends are consistent with variations in shell outflow velocity and \hi\ column density across the halos. However, some exceptions to these patterns are also noted.
Based on peak separation, red peak asymmetry, and residual trough flux, we identify five LAEs as strong LyC leaker candidates.

\end{abstract}

\section{Introduction}\label{sec:intro}

The \lya\ emission ($1215.67$ \AA) line of hydrogen atom is a key spectral feature that is used to detect and study high-redshift galaxies \citep[e.g.,][]{Dijkstra14, Hayes15, Ouchi-Ono20}. Due to their resonant nature, \lya\ photons scatter frequently in the interstellar,
circumgalactic, and even intergalactic medium (ISM, CGM, and IGM, respectively). This random walk leads to a wide diversity of spectral shapes of \lya\ that contain information about the kinematics and geometry of the gas present in ISM and CGM (see \citealt{Dijkstra19} for a review). The characteristic shape of \lya\ emission, making it easily recognisable, is typically a single red asymmetric line profile \citep[e.g.,][]{Shapley2003, Tapken2007}. However, over the past decade, double-peaked profiles of \lya, characterised by two asymmetric peaks, have been detected in both local galaxies \citep[e.g.,][]{Jaskot14, Henry15, Verhamme17, Orlitov18, Izotov18} and galaxies out to high redshifts \citep[i.e., $z = 2 - 7$;][]{Kulas2012, Yamada12, Matthee18, Hayes21, Meyer21, Songaila22, Erb23, Moya24, Vitte25}. Double-peaked profiles are also detected in Lyman alpha nebulae, or blobs \citep{Yang14, Vanzella17, Li22_blob}, suggesting complex radiative transfer effects and the presence of large-scale gas kinematics. In addition, a few triple-peaked profiles have also been reported \citep[e.g.,][]{RT17, Vanzella18, Vitte25}. The Multi-Unit Spectroscopic Explorer (MUSE) at the Very Large Telescope \citep[VLT;][]{Bacon10}, has revealed a vast population of \lya\ emitters (LAEs) with diverse spectral profiles at redshifts  $z = 2.9 - 6.7$ \citep[see][]{Herenz17, Bacon23, Vitte25}. Meanwhile, the Near Infrared Spectrograph aboard the James Webb Space Telescope (JWST/NIRSpec) is already extending the boundaries of \lya\ line observations to even higher redshifts \citep{Bunker24}.

Double-peaked \lya\ profiles typically arise due to scattering of photons from a static or expanding/infalling medium of neutral hydrogen cloud, which shifts their frequencies away from the line center and produces two peaks. In an outflowing medium, photons preferentially escape on the red side, producing a dominant red peak with a weaker blue bump. Additionally, \lya\ photons can backscatter off a receding outflow on the far side of the galaxy, which shifts their frequency out of resonance and allows them to escape without further scatterings \citep{Dijkstra2006, Verhamme06, Steidel10}, while multiple backscattering events can generate additional redshifted peaks and broad, extended red wings \citep{Verhamme06}.
A blueshifted peak usually appears when either the outflow has low opacity (e.g., lower column density or higher ionisation), enabling some photons to escape directly or when the photons scatter off of the neutral hydrogen (\hi) gas in
an inflowing CGM \citep{Zheng2002, Dijkstra2006}. Moreover, a dominant blueshifted peak over a redshifted peak arises when accretion/inflow dominates over outflow (see \citealt{Mukherjee23}; \citealt{Bolda24}, and references therein). Alternatively, double-peaked \lya\ profiles can also arise from two distinct, closely separated galaxies, each contributing its own \lya\ emission line peak with a velocity offset \citep{Cooke10}. In the local universe, the separation between the two \lya\ peaks is found to correlate with the escape of ionising Lyman-continuum (LyC) radiation, potentially due to low column density channel of \hi\ \citep[e.g.,][]{Verhamme17, Izotov18, Flury22b}, although this trend does not appear to hold at $z > 3$ \citep[see][]{Kerutt24}. Direct observations of the escaping LyC photons become nearly impossible at $z > 4$ due to significant attenuation by neutral hydrogen in the IGM (see \citealt{Ouchi-Ono20}; and references therein). Therefore, the double-peaked \lya\ profile emerges as an indirect
tracer of LyC escape, as well as the outflow and inflow kinematics that regulate the escape of both \lya\ and LyC photons.

The theory describing \lya\ radiative transfer has been studied for decades and for a range of gas geometries. A variety of powerful public codes have enabled numerical solutions to the \lya\ radiative transfer problem. The most widely used radiative transfer models typically describe outflows as spherical, expanding shells around a point-like \lya\-emitting region \citep[e.g.,][]{Verhamme2008, Verhamme15, Gronke15, GL22, GL25}. Despite their simple geometry, these models successfully reproduce various observed \lya\ profiles, including single and double-peaked features seen in Lyman break galaxies, LAEs, damped \lya\ systems, and Green Pea galaxies \citep[e.g.,][]{Verhamme2008, Vanzella10, Hashimoto15, Max_Gronke17, Yang17} and constrained properties of the scattering medium, such as \hi\ column density and shell velocity \citep[e.g.,][]{Verhamme15, Gronke17}. In general, a lower shell velocity enhances the blue peak, while a lower \hi\ column density reduces the peak separation \citep{Verhamme15}. However, physical parameters derived from the shell model often differ from those constrained by interstellar absorption and nebular emission lines \citep{Kulas2012, Orlitov18}. Recent studies have expanded \lya\ radiative transfer to multiphase, clumpy media, where cool \hi\ clumps are embedded in a hot, ionised inter-clump medium \citep[e.g.,][]{Neufeld91, Hansen2006, Laursen13, Gronke_Dijkstra2016, Li22}. 

\lya\ halos or extended \lya\ emission, extending several kiloparsecs around galaxies, trace CGM gas distribution, and kinematics through resonant scattering. Extended \lya\ halos are generally interpreted as arising from resonant scattering of \lya\ photons produced in central star-forming regions, rather than in-situ recombination at large radii. This is evident from local-universe studies (e.g. Lyman Alpha Reference Sample or LARS) showing \lya\ emission extending 2 to 4 times farther than \ha\ \citep{Hayes13, Ostlin14}, and by high-$z$ statistical analyses indicating that halo properties correlate with central \lya\ luminosity rather than halo mass \citep{Momose14, Kusakabe19}. However, detecting \lya\ halos at 
$z \gtrsim 2$ is challenging due to their faint emission and sensitivity limits. In recent years, IFU surveys such as MUSE and the Keck Cosmic Web Imager \citep[KCWI;][]{Martin10} have enabled the detailed study of individual halos around galaxies from $2 \lesssim z \lesssim 6$ \citep[see][]{Wisotzki16, Leclercq17, Erb18, Leclercq20, Erb23, Weldon24}. So far, MUSE studies have analysed the spectral properties across the halo around single-peaked LAEs, finding a correlation between line width and velocity shift of the peak, with broader emission
often tending to come from the outer regions of halo \citep{Claeyssens19, Leclercq20}. On the other hand, \cite{Erb23} and \cite{Weldon24} study variations in the peak ratio and separation of the
double-peaked \lya\ profile across the extended halos detected in KCWI, and find that peak separation decreases and the blue-to-red flux ratio increases toward the outskirts of the halo.

In this paper, we present a large sample of double-peaked \lya\ emitters identified through MUSE observations from the Middle Ages Galaxy Properties with Integral Field Spectroscopy (MAGPI) survey \citep{Foster21}, and provide a comprehensive analysis of their spectroscopic properties. In addition to characterising global spectral features, we conduct a spatially resolved analysis of double-peaked \lya\ profiles across ten extended halos at $z > 3$. To interpret these observations, we employ radiative transfer modeling to study the underlying gas kinematics and physical conditions driving \lya\ line formation.

The paper is structured as follows: In \S\ref{sec:obs}, we briefly describe the MAGPI observations, data reduction, and the method of identifying LAEs from datacubes. \S\ref{sec:sample selection} outlines our automated classification method for various \lya\ spectral shapes and the selection of double-peaked LAEs. In \S\ref{sec:forward}, we present a forward-modeling approach to fit the observed double-peaked profiles while accounting for the instrumental resolution. In \S\ref{sec:results}, we present both global and spatially resolved spectroscopic properties of our sample. \S\ref{sec:discussions} discusses these results in the context of radiative transfer models. We summarise our conclusions in \S\ref{sec:summary}. Throughout, we assume a flat $\Lambda$CDM cosmology with $H_0 = 70\, \mathrm{km}\,\mathrm{s}^{-1}\,\mathrm{Mpc}^{-1}$, $\Omega_{\mathrm{m}} = 0.3$, and $\Omega_{\Lambda} = 0.7$.



\section{Data}\label{sec:obs}

\subsection{Observations and data reduction}

The MAGPI survey\footnote{Based on observations obtained using MUSE instrument at VLT of the European Southern Observatory (ESO), Paranal, Chile (ESO program ID $1104.\mathrm{B}-0536$)} is an ongoing Large Program on the VLT/MUSE, targeting $56$ fields from the Galaxy and Mass Assembly \citep[GAMA;][]{Driver11} G12, G15 and G23 fields. MAGPI also includes archival observations of legacy fields Abell $370$ and Abell $2744$. The survey targets a total of $60$ primary galaxies with stellar masses $M_{*} > 7 \times10^{10}$\,$M_\odot$ and $\sim 100$ satellite galaxies with $M_{*} > 10^{9}$\,$M_\odot$. The primary objective of MAGPI is to perform a detailed spatially resolved spectroscopic analysis of stars and ionised gas within $0.25< z <0.35$ galaxies \citep[see][]{Foster21}. Data are taken using the MUSE Wide Field Mode ($1'\times1'$) with a spatial sampling rate of $0.2''$/pixel and the median Full Width at Half Maximum (FWHM) is $0.64''$, $0.6''$ and $0.55''$ in the $g$, $r$ and $i$ bands, respectively. Each field is observed in six observing blocks, each comprising 2$\times$1320\,s exposures, resulting in a total integration time of $4.4$\,h. The survey primarily employs the nominal mode (data are taken with the blue cut-off filter in place), providing a wavelength coverage ranging from $4700$\,\AA\ to $9350$\,\AA, with a dispersion of $1.25$\,\AA. Ground-layer adaptive optics (GLAO) is used to correct atmospheric seeing effects, resulting in a gap between $5805$\,\AA\ and $5965$\,\AA\ due to the GALACSI laser notch filter. The MUSE data reduction process is thoroughly described in \cite{Foster21}. 

\subsection{LAE identification}
\begin{figure}[ht!]
    \centering
    \includegraphics[width=0.98\linewidth]{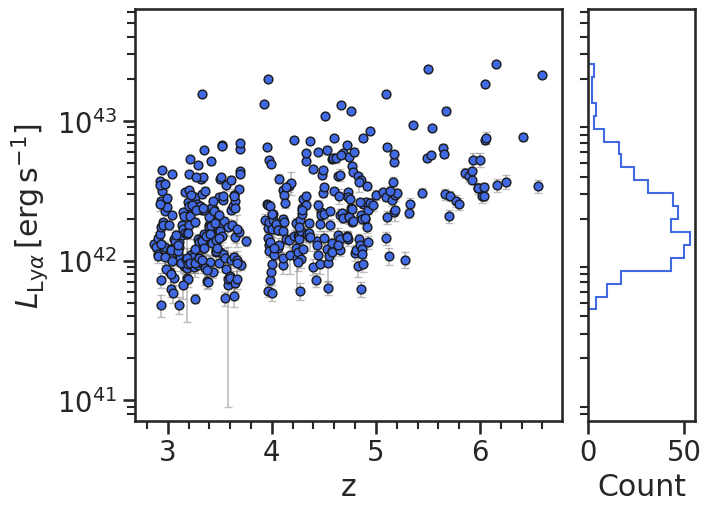}
    \caption{\lya\ luminosity as a function of spectroscopic redshift for 417 LAEs identified in first 35 MAGPI fields, spanning the redshift range $2.9 \lesssim z \lesssim 6.6$.}
    \label{fig:lum-z}
\end{figure}

The depth of the MAGPI data enables the detection of both foreground sources within the Local Universe and distant background sources, including LAEs in the redshift range $2.9 \lesssim z \lesssim 6.6$. To identify emission lines, we employ \texttt{LSDCat1.0}\footnote{\href{https://bitbucket.org/Knusper2000/lsdcat/}{https://bitbucket.org/Knusper2000/lsdcat/}} \citep{HW17}, a Python-based tool designed to detect faint emission lines in integral-field spectroscopic datacubes by correlating a matched filter with the data. A detection threshold of $\mathrm{SN}_{\mathrm{thresh}} = 7.0$ was adopted to minimize false positives. This choice is in line with the thresholds used in other MUSE-based LAE searches; for instance, the MUSE-WIDE survey applies a threshold of 5 – 6.4 for LAE detection \citep[see][]{Kerutt22}. The initial list of emission-line candidates is then passed through the \texttt{LAE\_Scanner}\footnote{\href{https://github.com/robbassett/LAE_scanner}{https://github.com/robbassett/LAE\_scanner}} tool, which filters out spurious and low-significance detections by enforcing criteria on spectral shape, substantially reducing the false-positive rate. For the remaining  candidates, segmentation maps were created using the \texttt{PROFOUND} R package \citep{Robotham18} and then global spectra are extracted. Redshifts are estimated using the \texttt{MARZ}\footnote{\href{https://github.com/Samreay/Marz}{https://github.com/Samreay/Marz}} software \citep{Hinton16}, which provides automated redshift identification based on spectral templates. During this step, all candidates undergo visual inspection to confirm the presence and shape of the \lya\ line, and to exclude contaminants such as low-redshift \oii\ emitters by checking whether the observed line profile matches the expected \oii\ $\lambda3726,3729$ doublet separation at lower redshifts. Following these steps, we identified 417 LAEs in the first 35 MAGPI fields. A circular aperture of $1.4''$ radius is used to compute spectroscopic properties, chosen to match the typical extent of LAEs while minimizing contamination from noisy spaxels. The redshift distribution of \lya\ luminosities is presented in Figure \ref{fig:lum-z}.

\section{Peak classification and sample selection}\label{sec:sample selection}

\begin{figure}[ht!]
    \centering
    \includegraphics[width=0.95\linewidth]{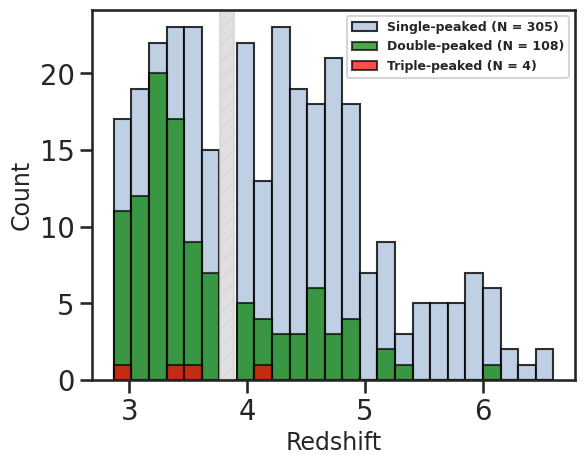}
    \caption{Redshift distribution of MAGPI LAEs classified by their \lya\ profile as outlined in \S \ref{sec:sample selection}: single-peaked (light-blue), double-peaked (green), and triple-peaked (red). The gray-shaded region ($z = 3.75$–$3.9$) marks a spectral gap due to the GALACSI laser notch filter, resulting in a lack of detected sources in this range.
    }
    \label{fig:hist}
\end{figure}

In this section, we outline the process of selecting double-peaked profiles from the MAGPI LAEs. Given that the observed \lya\ line profiles can show multiple peaks, we develop an automated method to systematically identify and characterise spectral features throughout the entire MAGPI LAE sample. The process involves the following steps: 


\begin{itemize}
    \item We first convert the observed wavelengths to vacuum wavelengths using the \texttt{pyasl} module from the \texttt{PyAstronomy} package. We then use a python package \texttt{pyplatefit}\footnote{\href{https://github.com/musevlt/pyplatefit}{https://github.com/musevlt/pyplatefit}} \citep{Bacon23} to fit and subtract any stellar continuum from each spectrum.  Finally, we convert these spectra from wavelength to velocity space, over a velocity window of $\pm 2000$ km/s, ensuring that this covers the entire \lya\ emission \citep{Kerutt22, Vitte25}.  

    \item We take the continuum-subtracted \lya\ spectra and obtain S/N spectra using the corresponding 1$\sigma$ uncertainties on flux densities. We smooth the S/N spectra with a 1-pixel moving average to reduce the impact of isolated noisy pixels. We then run \textit{find\_peaks} from \texttt{SciPy.signal} on the S/N spectra with a threshold of 1.5 to detect faint local maxima. For each local maxima, we locate the nearest local minima on either side and confirm whether at least two adjacent pixels in that window have S/N$\geq$1.5. If this holds, we mark those regions as significant with the maxima as possible peaks.

    \item We then fit a multi-asymmetric Gaussian function to the original spectra, focusing only on the regions around potential peaks and fixing the Gaussian mean to the corresponding peak velocity. The fitting function is defined as:
    \begin{equation}
    \begin{aligned}
    F \, (v) = \sum_{i=1}^{N} A_i \, \mathrm{exp} \left(-\, \frac{\Delta  v^{\, 2}_i}{2(a_{\mathrm{asym},i}\, \Delta v_i + w_i)^2}\right) 
    \end{aligned}
    \label{eqn1}
    \end{equation}
    where N is the number of potential peaks from the previous step, $A_i$ is the peak amplitude, $\Delta v_{i} = v - v_{0, i}$ is the velocity offset (in $\mathrm{km}\, \mathrm{s}^{-1}$), $a_{\mathrm{asym}, i}$ controls the asymmetry (or skewness), and $w_i$ is the line width (in $\mathrm{km}\, \mathrm{s}^{-1}$). We perform the fit using \texttt{curve\_fit} from the \texttt{scipy.optimize} package, incorporating the 1$\sigma$ flux uncertainties. During fitting, we impose two criteria: (i) the integrated S/N of each Gaussian must be $\geq$1.5; (ii) the FWHM must exceed the wavelength-dependent line spread function (LSF). We compute LSF for each MAGPI field using a second-order polynomial of the form: $\mathrm{FWHM} (\lambda) = A\, \lambda^{2} + B\, \lambda + C$ \citep[see e.g.,][]{Kerutt22}. This step results in the detection of up to $N=3$ significant peaks, classified as single, double, or triple-peaked profiles for $N=1,2,$ and $3$, respectively.

   \item 
   We then spatially verify those candidates using \lya\ narrow-band (NB) images. 
   For this, we use the MUSE Python Data Analysis Framework (\texttt{MPDAF}\footnote{\href{https://pypi.org/project/mpdaf/}{https://pypi.org/project/mpdaf/}}) package to generate NB images for each peak and compare centroids. In this process, we apply a signal-to-noise threshold of S/N = 1.5 to the NB images. A secondary peak is considered co-spatial if its NB centroid falls within a $0.6''$ radius aperture centered on the centroid of the primary peak (i.e., the peak with the highest integrated flux). This step reclassifies the peaks based on their spatial origin, as the two peaks in some double-peaked profiles arise from distinct spatial locations, indicating that the observed structure is not the result of radiative transfer effects. Therefore, we retain only the co-spatial peaks, resulting in a final classification of 108 double-peaked, 4 triple-peaked, and the remaining 306 classified as single-peaked profiles (Figure\,\ref{fig:hist}).   
\end{itemize}


 A comprehensive analysis of all MAGPI LAE profiles will be presented in Mukherjee et al. (in prep). Here, we focus exclusively on the 108 confirmed double-peaked LAEs, representing $\sim26$\% of the MAGPI LAEs. However, this fraction can be influenced by selection biases, including survey depth and the adopted S/N threshold. At $z<4$, we find a double-peaked fraction of $\sim37$\%, consistent with previous studies: $30$\% by \citet{Kulas2012}, $25$\% by \citet{Sobral18}, $32.9$\% for MUSE-Deep, and $33.8$\% for MUSE-Wide by \citet{Kerutt22}. As shown in Figure\ref{fig:hist}, the distribution of double-peaked LAEs peaks at $z \sim 3.2$ and rapidly drops beyond $z > 4$. This decline is expected due to the increasing neutral fraction of the IGM at higher redshifts \citep{Hayes21}. At $z > 4$, we find the double-peak fraction to be only $\sim 14$\%, consistent with the fraction reported for other MUSE observations \citep{Kerutt22} at similar redshifts.


\section{Forward modeling}\label{sec:forward}

\begin{figure}[ht!]
    \centering
    \includegraphics[width=0.98\linewidth]{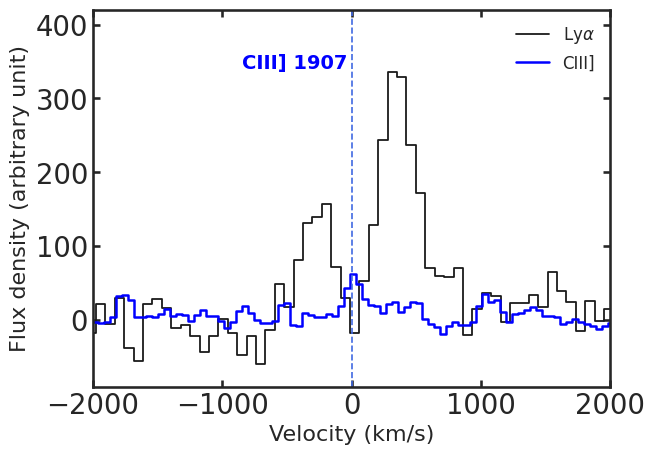}
    \caption{One example of rest-frame spectrum of a bright LAE (ID: 1201250296) with detected \ciii\ $\lambda 1907$ emission ($\sim 2.75 \, \sigma$) probing the systemic redshift $z_{\mathrm{sys}} = 3.2377$.}
    \label{fig:zsys}
\end{figure}

In this section, we introduce a forward modeling approach based on double-asymmetric Gaussian (DAG) functions to characterise the double-peaked profiles observed in our sample. The idea of a forward model is to take into account the MUSE instrumental broadening. To accurately infer the intrinsic properties of the \lya\ line profiles, it is crucial to account for this instrumental effect. Forward modeling provides a robust framework to do this. Before that, we first determine their systemic redshifts. For this, we apply the \texttt{pyplatefit} to search for \ciii\ $\lambda 1907$ \AA\ emission, a known proxy for systemic redshift in LAEs \citep{Verhamme18}. We detect \ciii\ emission with $>2\,\sigma$ significance in 14 sources at $z = 2.9$--$3.7$, and adopt their \ciii\ redshifts as systemic redshift ($z_{\mathrm{sys}}$). An example of this is presented in Figure \ref{fig:zsys}. Unfortunately, with the current MUSE data, we are limited in the precision we can achieve for these faint emission lines. Therefore, we caution that the reported systemic redshifts should be considered as indicative rather than definitive.
For LAEs without \ciii\ detection, we use the redshift of the absorption trough between the two \lya\ peaks as a proxy for $z_{\mathrm{sys}}$. 
In the forward modeling approach:

\begin{itemize}

    \item We begin with a double-asymmetric Gaussian  (i.e. N = 2 in Eq. \ref{eqn1}), with eight parameters in total. These parameters represent the intrinsic, pre-convolution state of the emission line, which is what the profile would look like without instrumental effects. 

    \item This model is then convolved with the wavelength dependent LSF, assumed to be Gaussian (with FWHM as mentioned in \S\ref{sec:sample selection}) to match the spectral resolution of the data. This convolution accounts for the spectral resolution of MUSE and models the blurring introduced by the instrument.

    \item Finally, we fit this convolved model to the observed double-peaked profiles. For this, we first used the  \texttt{curve\_fit} routine to perform a non-linear least squares fit to the observed spectrum. The purpose of this initial fit is to obtain reasonable estimates of the parameters. 

    \item The parameter estimates from curve\_fit is then used as the initial positions for the walkers in the Markov
    Chain Monte Carlo (MCMC) sampling, performed using the \texttt{emcee}\footnote{\href{https://github.com/dfm/emcee}{https://github.com/dfm/emcee}} package. MCMC explores the parameter space to derive the probability distribution. By setting \texttt{is\_weighted = True}, we force the emcee to directly sample the posterior distributions of the parameters with the correct weighting from $1\sigma$ uncertainty array, obtained from the variance cube. We use a sufficiently large chain length (steps = 10000) to ensure that the MCMC solution has converged.
\end{itemize}

\begin{figure*}[ht!]
    \centering
    {\includegraphics[width=18.5cm,height=5.5cm]{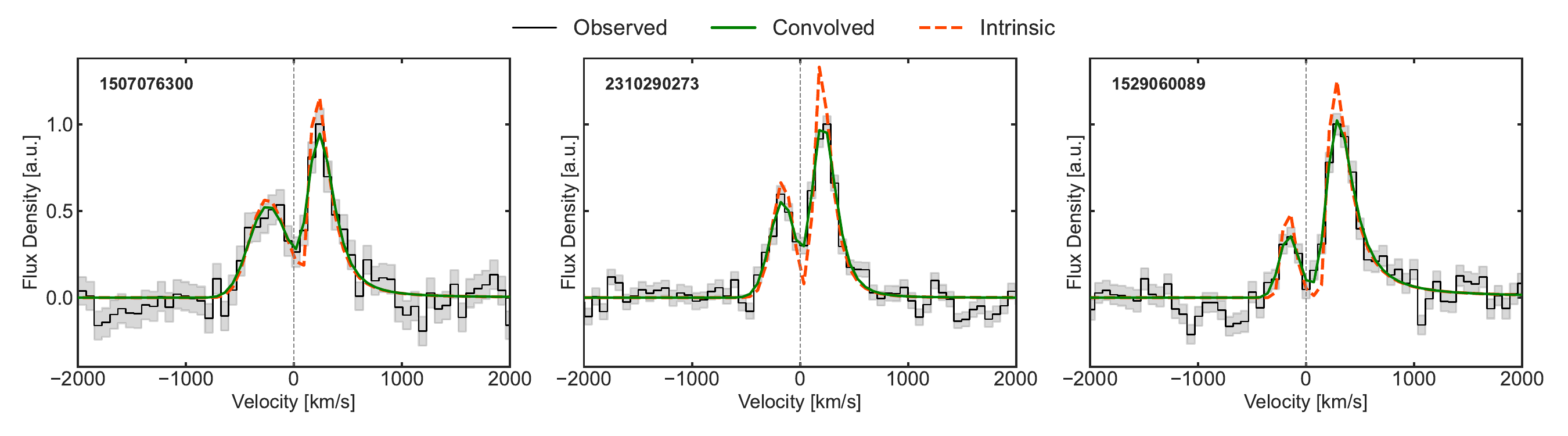}}
    \caption{Forward modeling of three representative double-peaked \lya\ spectra. The observed spectra are shown in black, with the best-fit forward-modeled profiles overplotted in green after convolution with the instrumental LSF. The intrinsic double asymmetric Gaussian models prior to convolution are shown in red. The corresponding MAGPI IDs are shown in the top left corner of each panel.}
    \label{fig:example_fit}
\end{figure*}

Figure \ref{fig:example_fit} shows how instrumental broadening affects the observed spectra. We present all the double-peaked profiles and their corresponding DAG fits in Figure~\ref{fig:DAG-fit} in the Appendix. The observed profiles are well described by these fits.


\section{Results}\label{sec:results}
We now present the analysis of the spectroscopic properties of the 108 confirmed double-peaked LAEs in our sample. In this section, we describe how key spectral measurements are obtained and outline the global and spatially resolved properties of these sources.

\subsection{Spectral Measurements} \label{sec:spectral_measurements}

We measure the following spectroscopic quantities from the intrinsic model parameters, reflecting true profiles before instrumental broadening:

\begin{itemize}
    \item Peak velocity separation ($\Delta_{\mathrm{peak}}$): Defined as the velocity difference between the red and blue peaks, $\Delta_{\mathrm{peak}} = V_{\mathrm{red}} - V_{\mathrm{blue}}$, where both velocities are measured with respect to the systemic redshift, in $\mathrm{km\,s^{-1}}$.

    \item Total \lya\ flux and luminosity: The total \lya\ flux is obtained by integrating the flux density under the area of the DAG. Flux uncertainties are drawn from the posterior distributions of the MCMC chains, providing parameter uncertainties that better reflect the full likelihood landscape. The \lya\ luminosity, $L_{\mathrm{Ly} \alpha}$, is derived from the total flux, and its corresponding errors.

    \item Integrated signal-to-noise ratio, (S/N)$_{\mathrm{total}}$: Calculated as the total integrated \lya\ flux divided by the flux uncertainty.

    \item Blue-to-total flux ratio (\bt): The relative strength of the blue peak is quantified as $F_{\mathrm{blue}} / F_{\mathrm{total}} = F_{\mathrm{blue}} / (F_{\mathrm{blue}} + F_{\mathrm{red}})$, where $F_{\mathrm{blue}}$ and $F_{\mathrm{red}}$ are the integrated fluxes of the blue and red peaks, respectively.

    \item FWHM of each peak: The line width of each peak is calculated using the fitting parameters \citep[see][]{Mukherjee24}:
    \begin{equation}
        \mathrm{FWHM}\, (\mathrm{km}\, \mathrm{s}^{-1}) = \frac{2\sqrt{2\,\ln(2)} \, w}{1 - 2\,\ln(2)\, a_{\mathrm{asym}}^2}
    \end{equation}
    where $w$ and $a_{\mathrm{asym}}$ are the width and asymmetry parameters of the fit, respectively. Uncertainties are derived from errors in the fit parameter returned by \texttt{emcee}.

    \item Flux density in the absorption trough ($F_{\mathrm{trough}}$): measured directly from the intrinsic DAG profile as the flux density at the minimum between the two peaks, with uncertainties estimated from MCMC sampling.

    \item \lya\ red peak asymmetry parameter, $A_{f}$: This is defined as follows:
    \begin{equation}
    A_{f} = \frac{\int_{\lambda_{\mathrm{red}}}^{\infty} f_{\lambda} d\lambda}{\int_{\lambda_{\mathrm{trough}}}^{\lambda_{\mathrm{red}}} f_{\lambda} d\lambda}
    \end{equation}
    where $\lambda_{\mathrm{red}}$ and $\lambda_{\mathrm{trough}}$ are the wavelengths of the red peak and absorption trough, respectively \citep{Rhoads2003, KG21}.   
\end{itemize}

\begin{figure}[ht!]
    \centering
    \includegraphics[width=0.98\linewidth]{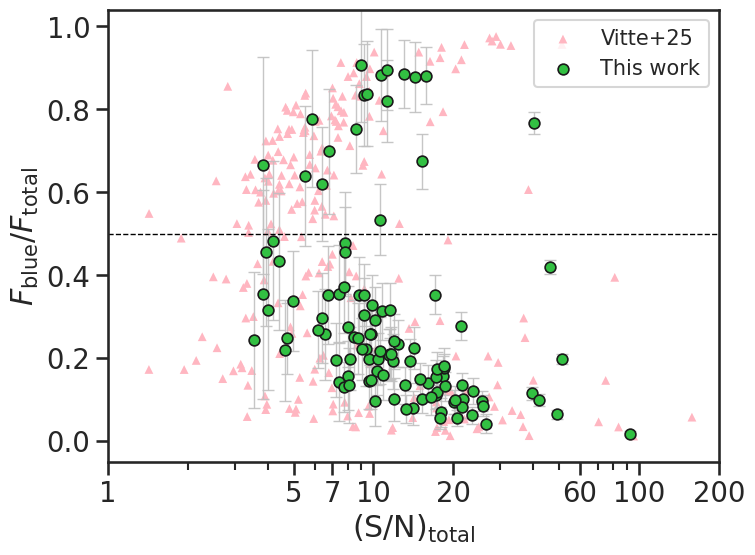}
    \caption{Blue-to-total flux ratio as a function of  the total S/N of the \lya\ line in our parent sample. Pink triangles are taken from \cite{Vitte25}. The horizontal black dashed line marks the division between blue- and red-dominated regions. }
    \label{fig:SNR}
\end{figure}

\subsection{Global \lya\ profiles}\label{sec:global}
This section describes the global properties of \lya\ emission measured from spatially integrated spectra. Full spectroscopic measurements are listed in Table \ref{t1} in the Appendix.

\subsubsection{Fraction of blue and red-dominated profiles}\label{sec:SN_ratio}
In our parent double-peaked LAE sample, the total integrated S/N of the \lya\ line ranges from 3.52 to 92.11.
We find $18/108 \, (\sim 17$\%) double-peaked profiles are blue-dominated (\bt\ $ > 0.5$). Three of these with MAGPI IDs 1534253170, 1207170349, and 2302106291 have systemic redshifts that confirm inflows or accretion-dominated sources. Of which two bright cases showing extended halos are reported in \cite{Mukherjee23}. The remaining 90 sources are red-dominated (\bt\ $ < 0.5$), indicating gas outflows. This blue-to-red fraction is in agreement with the cosmological zoom-in simulation \citep{Blaizot23} of a galaxy, where $<20$\% of the profiles are found to be blue-dominated. However, a recent study of LAEs from the MUSE Extremely Deep Field found a significantly higher fraction ($> 40$\%) \citep{Vitte25}. However, we note that numerous blue-dominated profiles lacking systemic redshift measurements could also arise from backscattering events \citep{Verhamme06}, where the observed smaller red bump originates from redshifted photons scattered off the far side of the outflowing medium. In particular, we observe several such cases at $z>4$. Without systemic redshifts, it remains uncertain whether these cases truly represent gas inflows or are instead manifestations of such backscattering effects.
Figure \ref{fig:SNR} shows the blue-to-toal flux ratio versus (S/N)$_{\mathrm{total}}$, including data from \cite{Vitte25}. We find that blue peak flux increases with decreasing (S/N)$_{\mathrm{total}}$, while extreme blue-dominated LAEs with \bt $> 0.8$ are only detected for (S/N)$_{\mathrm{total}} \gtrsim 7$.

\begin{table}[ht]
\centering
\caption{Summary of the double-peaked LAE sample used in this work.}
\label{tab:sample}
\begin{tabular}{lcc}
\hline\hline
Sample & Selection criterion & Number of sources \\
\hline
Parent sample      & All detected double peaks      & 108 \\
Conservative sample    & Both peaks with integrated S/N $\geq 3$        & 76  \\
\hline
\end{tabular}
\end{table}

For robust measurements, we adopt a conservative threshold of $(\mathrm{S}/\mathrm{N})_{\mathrm{blue}} \geq 3$ and $(\mathrm{S}/\mathrm{N})_{\mathrm{red}} \geq 3$ throughout the rest of the paper, resulting in a “Conservative sample” of 76 sources (see Table \ref{tab:sample}). Our conservative sample includes 9 blue-dominated LAEs.



\begin{figure}[ht!]
    \centering
    \includegraphics[width=0.999\linewidth]{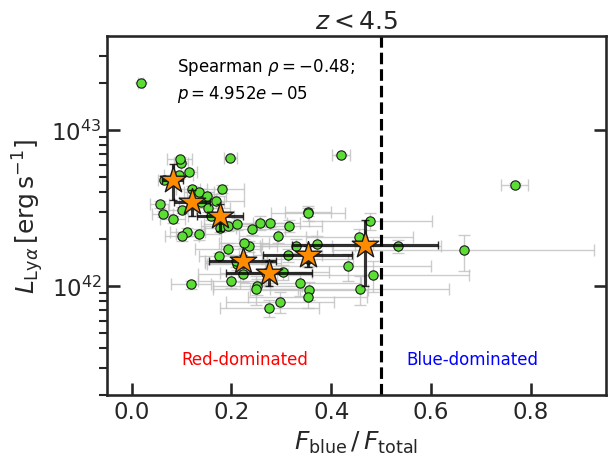}
    \caption{Observed \lya\ luminosity versus \bt\ for 64 LAEs in our conservative sample at $z < 4.5$, where the IGM has less effect. The vertical black dashed line separates the blue-dominated and red-dominated regions. Orange stars represent the median luminosity within each of the 7 equal-population bins (9 sources per bin), with the scatter in each bin indicated by black error bars. }
    \label{fig:lum}
\end{figure}

\begin{figure*}[ht!]
    \centering
    {\includegraphics[width=8.2cm,height=6.8cm]{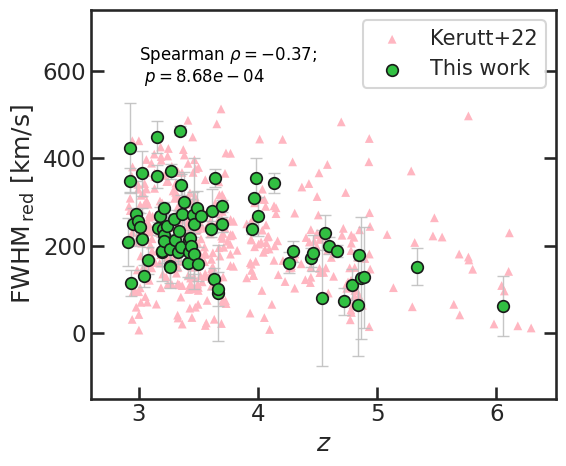}}
    {\includegraphics[width=9.4cm,height=6.86cm]{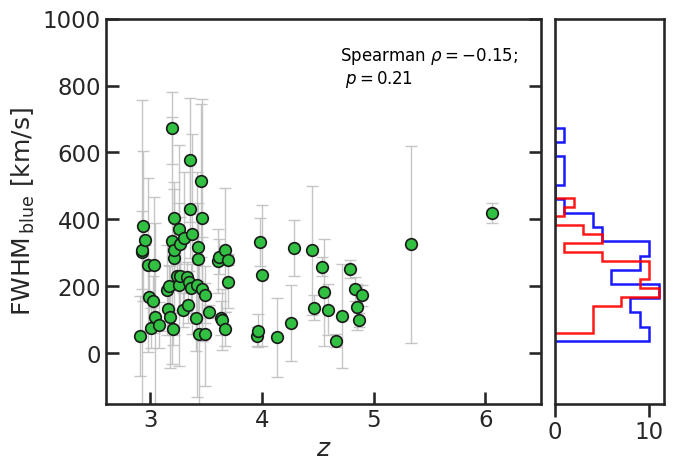}}
    \caption{Redshift distribution of \lya\ peak line widths in our conservative sample. The left panel shows the FWHM of the red peak ($\mathrm{FWHM}_{\mathrm{red}}$), while the right panel presents the FWHM of the blue peak ($\mathrm{FWHM}_{\mathrm{blue}}$), both obtained from DAG fitting. In the left panel, the pink triangle markers represent LAEs from the MUSE-Wide and MUSE-Deep surveys \citep{Kerutt22}.}
    \label{z_dependence}
\end{figure*} 

\subsubsection{\lya\ luminosity versus flux ratio}\label{sec:lum_ratio}

In Figure \ref{fig:lum}, we study the relationship between \lya\ luminosity and \bt\ for our conservative sample. For this, we take sources only up to $z < 4.5$. Sources at higher redshift ($z > 4.5$) are excluded, as they are strongly affected by IGM absorption and tend to be more luminous, which could bias the relation. We find a strong anticorrelation (Spearman $\rho = - 0.48$, p-value $= 4.95 \times 10^{-5}$) between \lya\ luminosity and \bt\ (Figure \ref{fig:lum}): Brighter \lya\ profiles show stronger red peaks, while fainter ones have more flux in the blue peak. We perform a quantile-based binning, taking the same number of sources within each \bt\ bin. We find that in the red-dominated region, the bin median luminosity decreases as \bt\ increases (see Figure \ref{fig:lum}). This trend is consistent with cosmological simulations \citep{Blaizot23} using mock \lya\ spectra. \citet{Blaizot23} identify this relation in a single simulated galaxy with a relatively constant stellar mass and star formation rate (SFR) during its evolutionary phase. Our results extend this trend to a broader population of double-peaked LAEs across a wide redshift range.

The relation flattens at lower luminosities and higher ratios.
The flattening is likely a selection effect rather than astrophysical. Since each \lya\ peak must have S/N $\geq 3$ in our conservative sample, galaxies with low total S/N (i.e., (S/N)$_{\mathrm{total}} \sim 5 - 6)$ can only be included if both peaks are of similar strength, driving \bt $\to 0.5$. This bias emerges near the luminosity where (S/N)$_{\mathrm{total}}$ drops below $\sim 6$ and may also contribute to the apparent slope at higher luminosities and lower ratios.


\subsubsection{Redshift distribution of line widths}\label{sec:z_dependence}
We find mean FWHMs of $\, 226.03 \pm  125.15\,\mathrm{km}\,\mathrm{s}^{-1} $ and $225.86 \pm 42.80 \,\mathrm{km}\,\mathrm{s}^{-1} $ for the blue and red peak, respectively. In Figure \ref{z_dependence}, we investigate the redshift dependence of the FWHM for both the red and blue peaks of the \lya\ line profile. We find that at $z < 4$, the red peak exhibits a wide range of line widths, including both narrow and broad profiles (see the left panel of Figure \ref{z_dependence}), indicating greater diversity in kinematic conditions or radiative transfer effects at these redshifts. In contrast, at $z \gtrsim 4$, the FWHM of the red peak shows a tendency to decrease with increasing redshift. We compare our data with that of \cite{Kerutt22} and notice a comprehensive trend for MUSE LAEs. This narrowing is not seen with high significance in the FWHM of the blue peak (right panel).  
Typically, at $z \gtrsim 4$, the abundance of \lya\ forest absorbers along the line of sight increases substantially, leading to enhanced attenuation of the \lya\ line profile. At even higher redshifts (i.e., $z \gtrsim 5$), the cumulative opacity of the IGM to \lya\ photons further increases due to the rising incidence of optically thick \hi\ absorbers. However, such absorptions cannot significantly attenuate the red peak without almost completely suppressing the blue peak. In our case, the persistence of a blue peak indicates that the absorptions in the IGM cannot be the primary cause of the red peak narrowing at high redshift.
Instead, the narrowing is more likely driven by intrinsic galaxy/CGM evolution. More luminous galaxies, which are generally observed at higher redshifts (Figure \ref{fig:lum-z}), are expected to exhibit broader intrinsic line widths prior to radiative transfer. The observed narrowing of the red peak, in contrast to the expected intrinsic broadening, further supports the interpretation that the effect arises from ISM/CGM evolution, such as lower outflow velocity dispersion, reduced \hi\ column density in outflows, or more ionised outflows at high redshift. The trend is influenced by limited statistics at $z > 5.5$,, where the likelihood of observing a double peak becomes nearly negligible due to the impact of the IGM.

We also note that the spectral resolution of MUSE improves with wavelength, from a line spread function (LSF) of approximately $170 \, \mathrm{km}\,\mathrm{s}^{-1}$ at $4700$ Å to about $90 \, \mathrm{km}\,\mathrm{s}^{-1}$ at $9350$ Å. This enhances resolution at longer wavelengths and makes it easier to resolve and detect narrow features, particularly the blue peak, at higher redshifts. However, despite this advantage, the consistent suppression of the blue peak's line width across redshift further underscores the strong and persistent impact of IGM attenuation on the blue side of \lya\ emission.

\begin{figure}[ht!]
    \centering
    \includegraphics[width=0.99\linewidth]{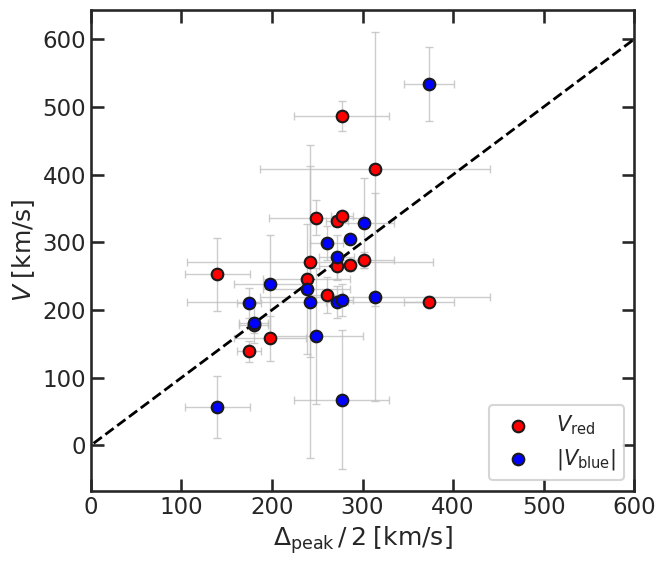}
    \caption{Velocities of \lya\ blue and red peak are plotted against half of the peak separation ($\Delta_{\mathrm{peak}}/2$) for LAEs with known systemic redshifts in our sample. The black dashed line shows the one-to-one relation.
    }
    \label{sep}
\end{figure}

\subsubsection{Correlations among velocities, peak separation and line width}\label{4.2.4}
Our conservative sample spans \lya\ peak separations of $\Delta_{\mathrm{peak}} = 223 - 1008\, \mathrm{km}\,\mathrm{s}^{-1}$, with a mean of $498.98 \pm 73.12 \, \mathrm{km}\,\mathrm{s}^{-1} $. This is consistent with previous findings from MUSE-Wide and MUSE-Deep LAEs \citep[$481 \pm 244 \, \mathrm{km}\,\mathrm{s}^{-1} $;][]{Kerutt22}, the MUSE Extremely Deep Field \citep[$534 \pm 28 \,\mathrm{km}\,\mathrm{s}^{-1}$;][]{Vitte25}, and $z \sim 2.2$ LAEs \citep[$500 \pm 56 \,\mathrm{km}\,\mathrm{s}^{-1}$;][]{Hashimoto15}. 
Our sample remains above the minimum measurable threshold of $150\, \mathrm{km}\,\mathrm{s}^{-1}$ reported by \citet{Vitte25}.

For LAEs with known systemic redshifts, we display the blue and red peak velocities as a function of half the peak separation ($\Delta_{\mathrm{peak}}/2$) in Figure~\ref{sep}, and we notice a strong correlation close to the one-to-one relation. This indicates that double-peaked profiles are typically symmetrically centered around the systemic velocity.
\citet{Pahl24} report that the ionising photon escape fraction, \fesc\, correlates more strongly with both $\Delta_{\mathrm{peak}}$ and $V_{\mathrm{blue}}$ than with $V_{\mathrm{red}}$ at $z \sim 0.3$. This suggests that the visibility of the blue peak in LAEs is closely linked to the escape of ionising photons, potentially indicating that similar neutral-phase gas dynamics govern both the escape of the blue \lya\ peak and the ionising photons. 

\begin{figure}[ht!]
    \centering
    \includegraphics[width=0.99\linewidth]{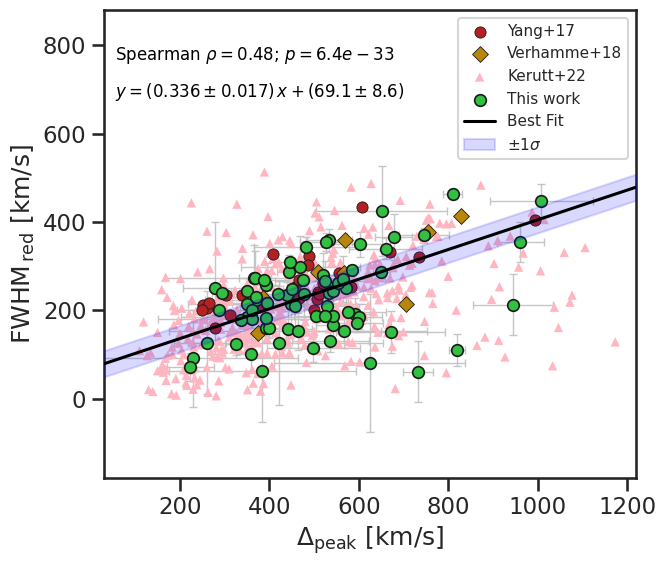}
    \caption{FWHM of the main (red) peak as a function of double-peak separation of \lya\ . Red circles represent $z \sim 0.3$ Green Pea galaxies compiled by \citet{Yang17}. Yellow diamonds denote $z = 3-4$ LAEs from \citet{Verhamme18}, using objects with peak separations listed in their Table~1., and pink triangles correspond to LAEs at $z \sim 2.9 - 6.6$ from the MUSE-Wide and MUSE-Deep surveys \citep{Kerutt22}. The best-fit to the combined data is shown as a black solid line, with the corresponding $1\sigma$ confidence bounds indicated by blue dashed lines. The Spearman correlation coefficients and the equation of the best-fit line are displayed in the top left corner. 
    }
    \label{fig:FW_sep}
\end{figure}

We find a strong correlation between $\mathrm{FWHM}_{\mathrm{red}}$ and $\Delta_{\mathrm{peak}}$ (see Figure \ref{fig:FW_sep}), indicating that broader red components are associated with larger peak separations. This trend is consistent with both observations \citep{Yang17, Verhamme18, Kerutt22} and radiative transfer models \citep{Verhamme15}. To carry out a linear regression, we merge our dataset with those of \cite{Kerutt22}, \cite{Yang17}, and \cite{Verhamme18}, and use \texttt{LtsFit}\footnote{\href{https://pypi.org/project/ltsfit/}{https://pypi.org/project/ltsfit/}} \citep{Cappellari13}, which accounts for uncertainties in both variables and incorporates intrinsic scatter in the data. The combined data show Spearman correlation coefficients of p-value = $6.4\, \times 10^{-33}, \rho = 0.48$.
The intrinsic scatter estimated from the fit is $29.6 \, \mathrm{km}\,\mathrm{s}^{-1}$. The resulting best-fit, shown by the solid black line in Figure \ref{fig:FW_sep}, is given by
\begin{equation}
    \mathrm{FWHM}_{\mathrm{red}} = (0.336 \pm 0.017) \Delta_{\mathrm{peak}} + (69.1 \pm 8.6)
\end{equation}


\begin{figure*}[ht!]
    \centering
    {\includegraphics[width=8.9cm,height=7.5cm]{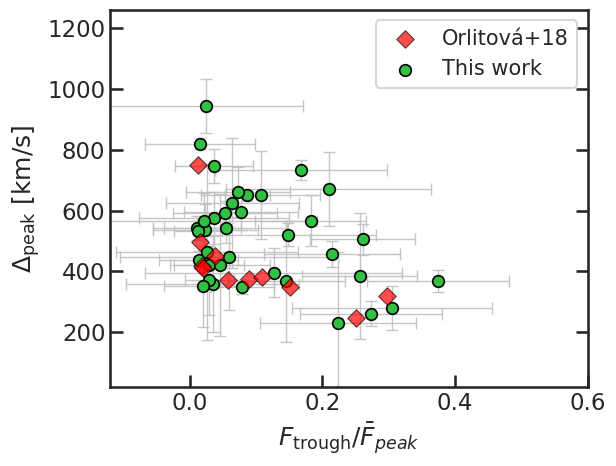}}
    {\includegraphics[width=8.9cm,height=7.5cm]{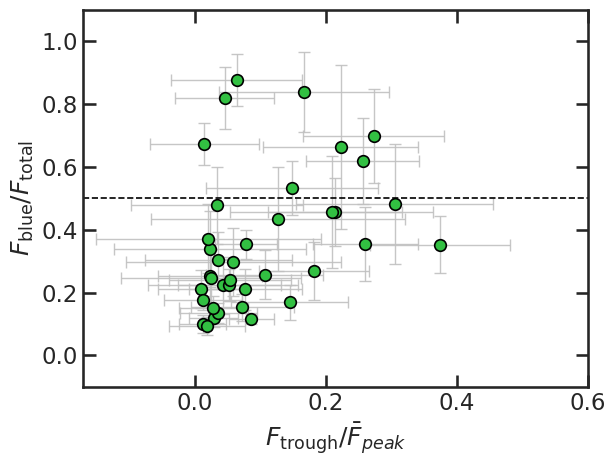}}
    \caption{\lya\ double peak separation (left panel) and blue-to-total flux ratio (right panel) versus normalised trough flux density $F_{\mathrm{trough}}/ \bar{F}_{\mathrm{peak}}$ for the conservative sample. Red diamonds are data points from \cite{Orlitov18}. Only sources with $F_{\mathrm{trough}}/ \bar{F}_{\mathrm{peak}} > 0.009$, corresponding to the lowest value reported in \cite{Orlitov18}, are presented. In the right panel, the horizontal black dashed line indicates the boundary between blue- and red-dominated regions. }
    \label{fig:sep_trough}
\end{figure*}

\begin{figure*}[h!]
\centering
    {\includegraphics[width=4.4cm,height=3.9cm]{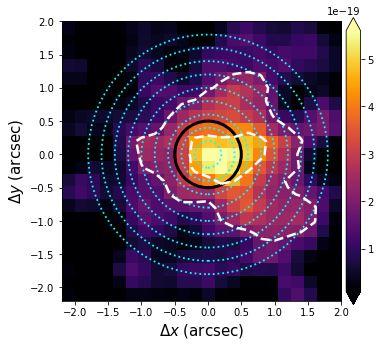}}
    {\includegraphics[width=4.6cm,height=3.9cm]{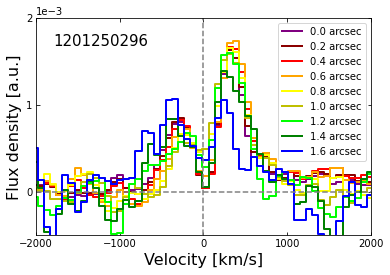}}
    {\includegraphics[width=4.4cm,height=3.8cm]{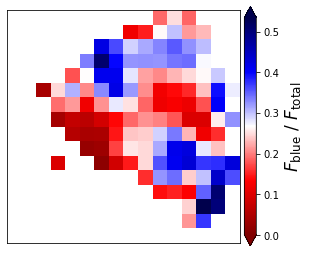}}
    {\includegraphics[width=4.4cm,height=3.8cm]{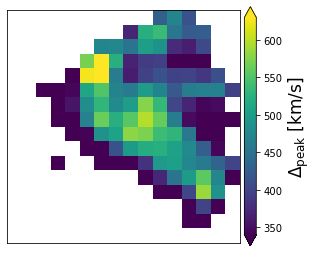}}
    \medskip
    \caption{Spatially resolved properties of one extended \lya\ halo (ID: 1201250296) in our sample. 1st panel: \lya\ pseudo narrow-band image; Coordinate is centred on the brightest pixel. Shown contours in white represent 2 and 4 $\sigma$ significance levels. The black circle denotes the location of stellar continuum detected in the MUSE white light image. The regions of spectral extraction using circular annular binning (as as described in \S \ref{5.2}) are shown as cyan circles. 2nd panel: Spectra from annular bins. 3rd and 4th panels: Pixel-by-pixel maps of the blue-to-total flux ratio ($F_{\mathrm{blue}}\, / F_{\mathrm{total}}$) and peak separation, respectively, shown only for regions with $\geq 2 \, \sigma$ significance and zoomed to this area, and therefore not on the same spatial scale as the narrow-band image. These two maps are  showing azimuthal variations in flux ratio and peak separation across the halo.}
    \label{fig:SR_example}
\end{figure*}

This result implies that the redshifted \lya\ emission broadens as it escapes further from the systemic velocity. This behavior has been interpreted via radiative transfer model using homogeneous shell models \citep{Verhamme18}, simulations considering anisotropic or bipolar outflows and inhomogeneous gas distributions \citep{Zheng14}, and multiphase clumpy media models \citep{Gronke_Dijkstra2016}. Together, these models predict a robust $V_{\mathrm{red}}$–$\mathrm{FWHM}_{\mathrm{red}}$ correlation independent of outflow geometry or kinematics. 

\subsubsection{Trough flux density distribution}\label{sec:4.2.5}
We find that several LAEs in our sample show positive residual flux in the absorption trough, $F_{\mathrm{trough}}$, between the two peaks. The ratio of $F_{\mathrm{trough}}$ and the mean flux of the peak maxima, $\bar{F}_{\mathrm{peak}}$, serves as a diagnostic to distinguish clumpy from homogeneous shell models, showing different distributions in each \citep{Gronke_Dijkstra2016}. Throughout the rest of this paper, we refer to this  ratio\footnote{The quantity $F_{\mathrm{trough}}/ \bar{F}_{\mathrm{peak}}$ used here is not identical to the central escape fraction $f_{\mathrm{cen }}$ defined in \cite{Naidu22}. However, the physical motivation is similar: both quantities parameterise the relative amount of \lya\ flux transmitted near line centre, and are therefore conceptually equivalent in tracing central \lya\ escape.} ($F_{\mathrm{trough}}/ \bar{F}_{\mathrm{peak}}$) as the normalised trough flux density. Our result shows an anticorrelation between this ratio and peak separation (Figure \ref{fig:sep_trough}, left panel), consistent with \citet{Orlitov18} for the green pea galaxies at $z \sim 0.2 - 0.3$. Such a correlation is absent in the clumpy model \citep{Gronke_Dijkstra2016}. While the clumpy model predicts $F_{\mathrm{trough}}/ \bar{F}_{\mathrm{peak}}$ up to $\sim 0.8$, our sample spans only between $0$ – $0.4$.
 We also find large scatter in peak separation for $F_{\mathrm{trough}}/ \bar{F}_{\mathrm{peak}} \sim 0 $, and a decrease to $\lesssim 500\, \mathrm{km}\, \mathrm{s}^{-1}$ for $F_{\mathrm{trough}}/ \bar{F}_{\mathrm{peak}} \gtrsim 0.2 $, supporting the predictions from the homogeneous shell model \citep{Orlitov18}.

In Figure \ref{fig:sep_trough} (right panel), we plot \bt\ against $F_{\mathrm{trough}}/ \bar{F}_{\mathrm{peak}}$, revealing distinct distributions in blue- and red-dominated regions. In red-dominated regions, the trough flux rises with increasing blue peak flux. In contrast, strong blue-dominated LAEs (\bt\ $> 0.8$) show low trough flux ( $F_{\mathrm{trough}}/ \bar{F}_{\mathrm{peak}} < 0.2$). This is likely due to the S/N of the \lya\ line, given that the blue-to-total flux ratio shows a strong dependence on (S/N)$_{\mathrm{total}}$, as discussed in \S\ref{sec:SN_ratio}.




\begin{figure*}[ht!]
\centering
    {\includegraphics[width=18cm,height=12cm]{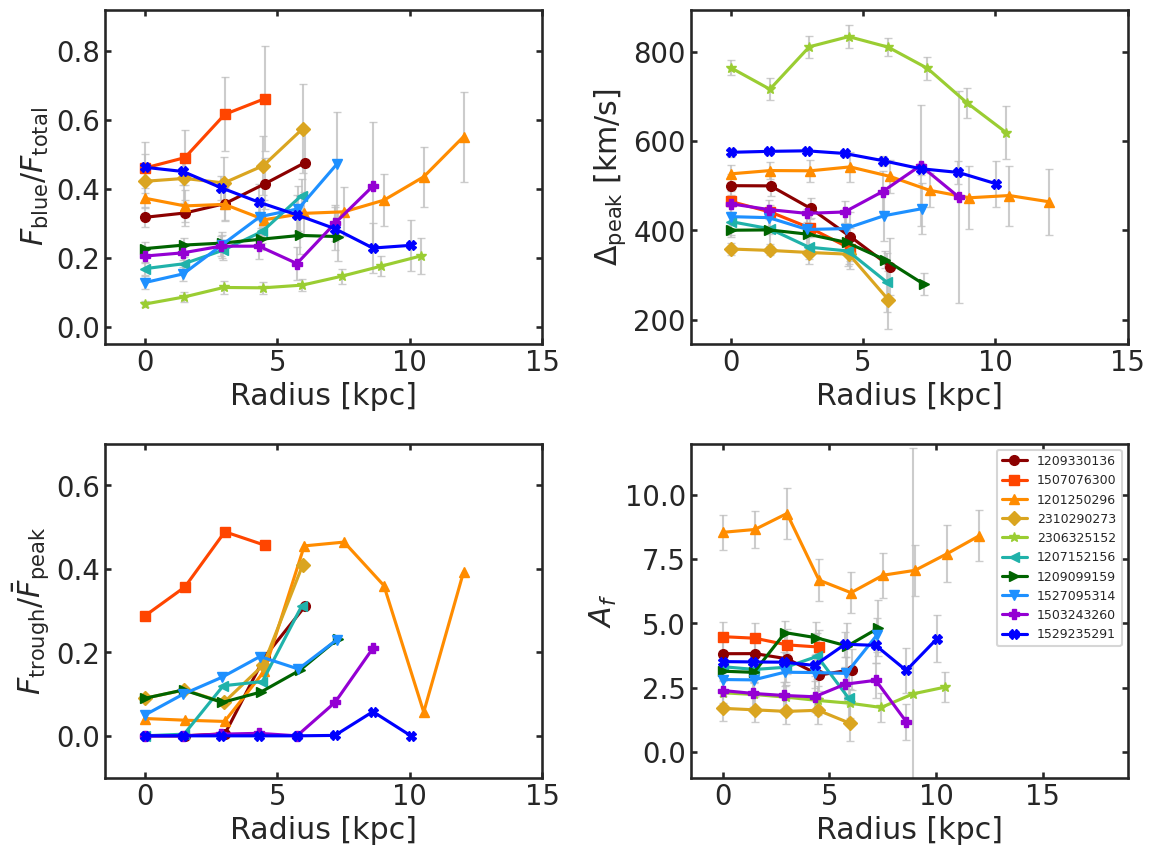}}
\caption{Shown are results from line profile measurements of the spatially resolved spectra extracted from the annular regions described in \S\ref{5.2}. (top row) Left panel: blue-to-total flux ratio ($F_{\mathrm{blue}}\, / F_{\mathrm{total}}$) as a function of radius (kpc), Right panel: \lya\ peak separation versus radius; (bottom row) Left panel: normalised trough flux density ($F_{\mathrm{trough}}/ \bar{F}_{\mathrm{peak}}$) versus radius, Right panel: \lya\ red peak asymmetry ($A_{f}$) versus radius. MAGPI IDs corresponding to each galaxy are indicated in the Right panel of bottom row.
}
\label{fig:resolved}
\end{figure*}

\subsection{Spatially resolved \lya\ profiles}\label{5.2}
Integral field spectroscopy has enabled spatially resolved studies of LAEs. To investigate general trends and analyse the spectral properties of double-peaked LAEs as a function of radius, we employ a method similar to \citet{Erb23}. Among our 108 LAEs, we select 10 bright LAEs that show double-peaked profiles throughout their spatial extent with good S/N and exhibit clear spatial variations in their spectral profiles. They show extended \lya\ halos in the continuum-subtracted narrowband images. While our selection ensures robust spatially resolved measurements, we note that double-peaked morphology does not necessarily persist at all radii in general. For example, \citet{Weldon24} report a halo with a single peak in the central region but double peaks in the inner and outer halo.

Annular spectra are created by binning spaxels radially. The high S/N of the \lya\ halos enables reliable spatial binning. Starting from the central spaxel with the highest surface brightness, we include all spaxels with S/N $> 2$ in the continuum-subtracted \lya\ narrowband images. Each halo is binned in radial increments corresponding to a single spaxel ($0.2''$). The spectra within each radial bin are summed to construct \lya\ profiles for different radii. We normalise these spectra by the total \lya\ flux to clearly witness the variations in both peaks. We then fit a double-asymmetric Gaussian to these spectra by forward modeling, as in \S\ref{sec:forward}. In addition to analysing spatially averaged profiles, we also create pixel-by-pixel maps of the blue-to-total flux ratio and peak separation across the halos to investigate their azimuthal variations.
Figure \ref{fig:SR_example} shows an example of an extended \lya\ halo in our sample, where the pseudo-NB image displays the extended emission. The spatially-resolved properties for all ten halos at $z = 3.2 - 3.7$ are shown in Figure \ref{SR_properties} in the Appendix.


Figure \ref{fig:resolved} presents the spatial variations of the spectroscopic quantities as a function of radius in kpc. Halo radii span $4.6$–$12$ kpc. In 9 of 10 halos, $F_{\mathrm{blue}} / F_{\mathrm{total}}$ increases with radius, consistent with $z \sim 2$ double-peaked \lya\ studies \citep{Erb23, Weldon24}, likely due to line-of-sight effects in the outer halo showing more flux in the blue peak. In addition, peak separation decreases with radius, in agreement with the findings of \cite{Erb23} and \cite{Weldon24}. However, we find that these spatial variations are not always circularly symmetric. Instead, significant azimuthal differences are present in some cases (see the example given in Figure \ref{fig:SR_example}, and also see Figure \ref{SR_properties} in the Appendix), with certain regions within the halo exhibiting enhanced blue flux and reduced peak separation, rather than these features being uniformly distributed from the core to the outskirts. As a result, this azimuthal structure also somewhat influences the interpretations drawn from spatially averaged profiles based on annular regions. Azimuthal variations in the peak ratio suggest velocity asymmetries and nonradial gas motions at large radii. We also observe that the normalised trough flux density generally shows an increasing trend with radius, while \lya\ red peak asymmetry shows a mild anticorrelation for some of them, but no strong trend is observed.
Generally, beyond $\sim4$ kpc, the blue flux and the flux in the absorption trough rise,  while the peak separation falls from center to outward, though the small sample limits physical constraints on this scale. 


We also see some exceptions in these trends.
Interestingly, one halo (ID: 1529235291) shows the opposite trend: a decreasing flux ratio with radius, while peak separation still decreases. Similar behavior is seen in blue-dominated halos \citep{Mukherjee23, Bolda24}, where the central regions with the highest surface brightness are blue-dominated and the outskirts show a stronger red peak. We also notice a moderate decrease in normalised trough flux density with radius for this source. Additionally, the halo with ID: 1503243260 shows a slight increase in peak separation with radius.
We explore these trends in more detail in the next section in the context of spatially resolved radiative transfer modeling.



\section{Discussions}\label{sec:discussions}
In this section, we discuss the spectroscopic properties in the context of \lya\ radiative transfer simulation at both global and local scales.

\subsection{Radiative transfer modeling}\label{sec:RTmodeling}

Studies on \lya\ radiative transfer through clumpy shells have shown that the \lya\ spectral profiles can be more complex than those from homogeneous shells. In this work, we adopt the simple and widely used homogeneous shell model. We use the publicly available radiative transfer code \texttt{\texttt{zELDA}}\footnote{\href{https://github.com/sidgurun/Lya_zELDA/}{https://github.com/sidgurun/Lya\_zELDA}} \citep{GL22} to model both spatially integrated and resolved double-peaked \lya\ profiles. \texttt{zELDA} is based on radiative transfer code \texttt{FLaREON} \citep{GL19}. This code employs a thin-shell geometry for the gas distribution surrounding the \lya\ emitting region, assuming an isothermal, homogeneous, spherical shell of neutral hydrogen at $10^4$ K, characterised by a uniform radial bulk velocity, $V_{\mathrm{shell}}$. Importantly, \texttt{zELDA} accommodates both inflow ($V_{\mathrm{shell}} < 0$) and outflow ($V_{\mathrm{shell}} > 0$) scenarios, with emergent spectra from inflow is identical to those of a shell with an outflow velocity, but mirrored around the zero \lya\ velocity (systemic). This framework enables an effective analysis of the gas dynamics within the ISM and CGM of the studied LAEs. The model incorporates five free parameters: shell velocity ($V_{\mathrm{shell}}$ in km $\mathrm{s}^{-1}$), neutral hydrogen column density ($\log(N_{HI})$ in cm$^{-2}$), dust optical depth ($\tau$), intrinsic equivalent width ($\log(EW)_{\mathrm{int}}$ in \AA), and intrinsic line width ($W_{\mathrm{int}}$ in \AA). \lya\ and continuum photons are generated from the central source with an intrinsic width of $W_{\mathrm{int}}$ and an intrinsic equivalent width of $\log(EW)_{\mathrm{int}}$ and propagate through the \hi\ shell, undergoing absorption or resonant scattering while traveling.

{\texttt{zELDA} performs forward modeling that accounts for the instrument’s resolution and pixelisation, and employs an MCMC approach to fit the observed line profiles. The S/N of the spectrum is also incorporated. To estimate uncertainties, the shell parameters and redshift are derived from the observed spectrum after applying noise perturbations 1000 times. The best-fit models effectively reproduce key features of global double-peaked \lya\ profiles. 
The best-fit \hi\ column densities range from log$(N_{\mathrm{HI}}) = 18.36 - 21.12 \, \mathrm{cm}^{-2}$, which is consistent with the typical range observed in previous LAE studies.


\subsection{Global spectroscopic properties and $N_{\mathrm{HI}}$ dependence} 

Radiative transfer modeling of \lya\ emission, using both simplified shell geometries and more complex clumpy configurations, has successfully reproduced observed correlations between spectral features and physical model parameters. For example, a strong correlation has been found between the separation of the \lya\ peaks and $N_{\mathrm{HI}}$ in the ISM and CGM of galaxies \citep[see][]{Verhamme15, Verhamme18, Li22}. A lower $N_{\mathrm{HI}}$ allows \lya\ photons to escape with fewer scatterings, resulting in narrower lines and leading to a smaller peak separation. Conversely, at very high redshifts ($z > 6$), LAEs within larger ionised bubbles exhibit broader red peaks \citep[e.g.,][]{Mukherjee24}. Additionally, the removal of neutral gas at zero velocity due to early photoionisation facilitates the escape of \lya\ photons on the blue side \citep{Hayes23}. On the other hand, the strength of the blue peak tends to decrease with increasing shell outflow velocity \citep[e.g.,][]{Verhamme15, Li22, Erb23}. 

\begin{figure}[ht!]    
    \includegraphics[width=0.99\linewidth]{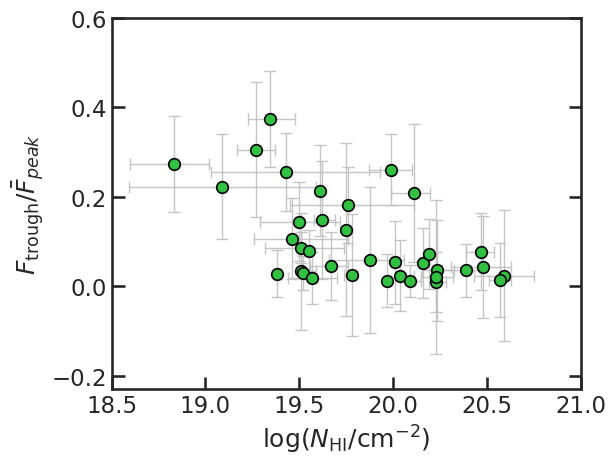}
    \caption{Normaslied trough flux density $F_{\mathrm{trough}}/ \bar{F}_{\mathrm{peak}}$ versus global neutral hydrogen column density ($N_{\mathrm{HI}}$) for sources with $F_{\mathrm{trough}}/ \bar{F}_{\mathrm{peak}} > 0.009$ (similar to Figure \ref{fig:sep_trough}). }
    \label{NH_dependence}
\end{figure}

\begin{figure*}[ht!]
    \centering
    
    \subfloat[]{%
        \includegraphics[width=4.3cm,height=3.3cm,clip]{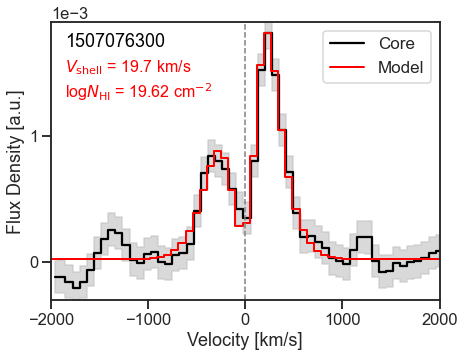}
        \includegraphics[width=4.3cm,height=3.3cm]{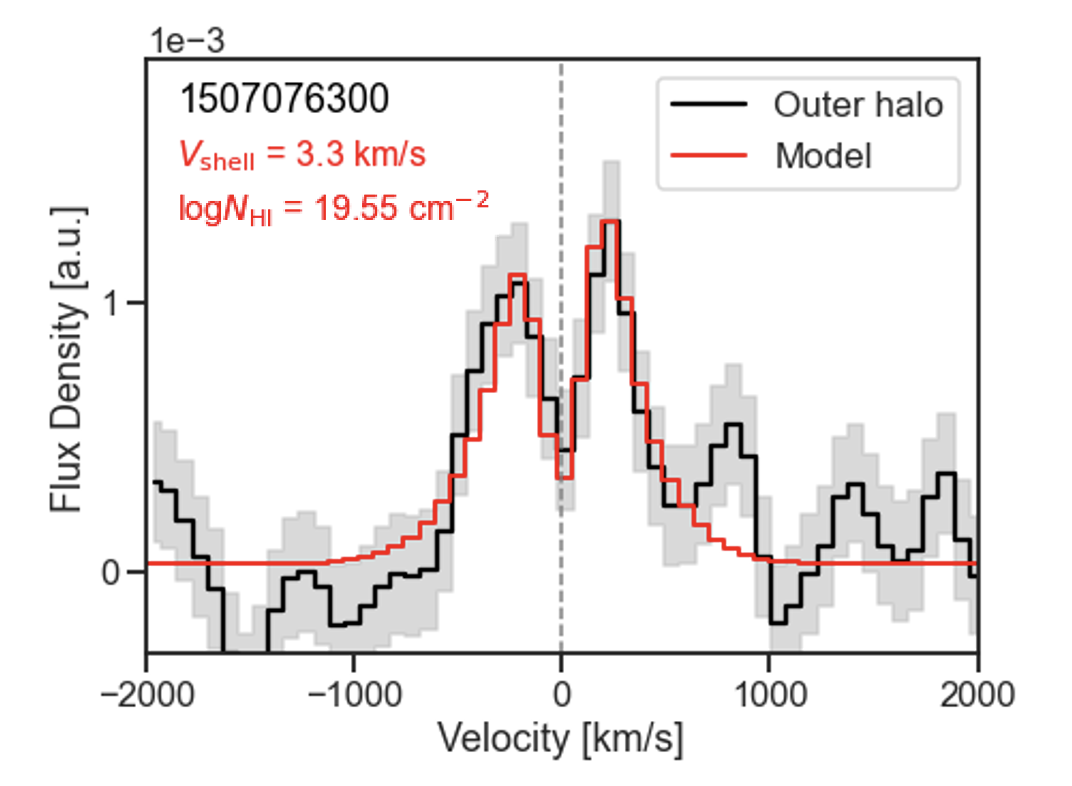}
    }\hspace{0.5cm}
    \subfloat[]{%
        \includegraphics[width=4.3cm,height=3.3cm,clip]{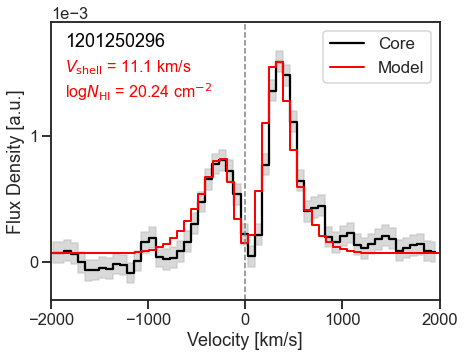}
        \includegraphics[width=4.3cm,height=3.3cm,clip]{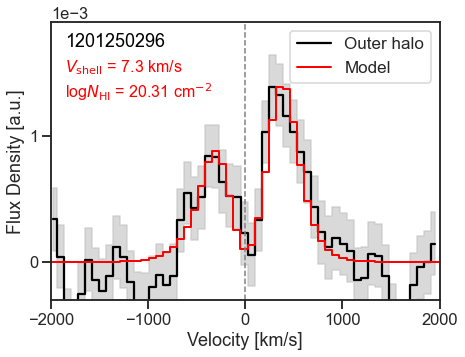}
    }
    
    \medskip
    
    \subfloat[]{%
        \includegraphics[width=4.3cm,height=3.3cm]{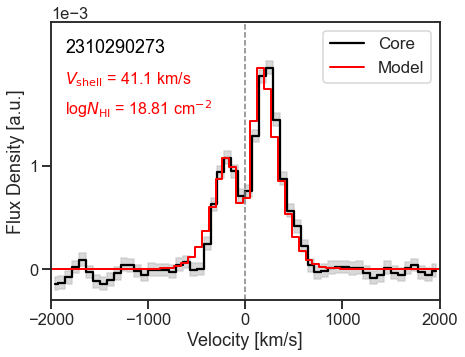}
        \includegraphics[width=4.3cm,height=3.3cm]{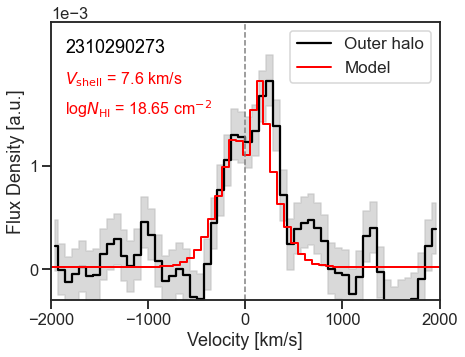}
    }\hspace{0.5cm}
    \subfloat[]{%
        \includegraphics[width=4.3cm,height=3.3cm]{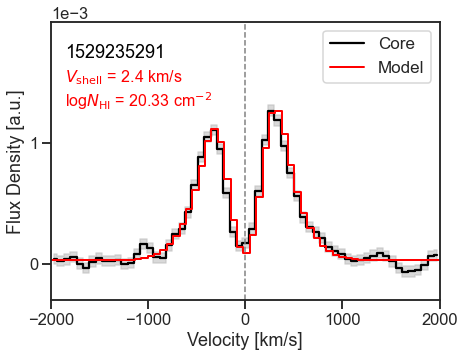}
        \includegraphics[width=4.3cm,height=3.3cm,clip]{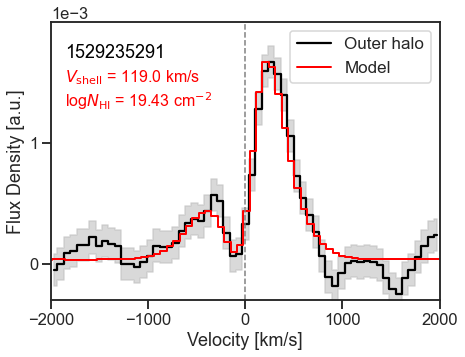}
    }
    
    \vspace{-0.3cm}
    \caption{Examples of spatially resolved radiative transfer modeling of \lya\ across four bright halos are shown. Observed spectra (black) and best-fit \texttt{zELDA} models (red) are presented for the `Core' and `Outer halo', with best-fit shell outflow velocities and \hi\ column densities indicated in the top-left corner of each panel.}
    \label{fig:resolved_examples}
\end{figure*}

The IGM predominantly absorbs the blue peak, diminishing its line flux, while the red peak is shifted out of the resonance frequency. Stacked \lya\ spectra reveal a decreasing fraction of blue-side flux with increasing redshift \citep{Hayes21, Kramarenko24}, making double-peaked structures challenging to observe at high redshifts \citep{MG20}. Consequently, the number of double-peaked LAEs decreases rapidly with redshift (see Figure~\ref{fig:hist}), as the increasingly neutral IGM attenuates the blue peak, leading to single-peak detections. Ionised bubbles around high-redshift LAEs may allow blue peaks with small separations to shift out of resonance before encountering the IGM. Therefore, we observe minimal redshift effects on $\mathrm{FWHM}_{\mathrm{blue}}$. \citet{Kerutt22} suggest that smaller peak separations correspond to smaller ionised bubbles necessary for the blue bump to shift sufficiently to survive IGM attenuation. More realistic radiative transfer models are required to further constrain these properties.


In Figure \ref{NH_dependence}, the residual trough flux density is shown as a function of the global \hi\ column density. In general, we find that LAEs with relatively lower column densities typically show a higher trough flux. The higher the column density, the more scattering and absorption \lya\ photons experience near the line center. This pushes \lya\ photons further into the red and blue wings before they can escape. This results in greater suppression of the central flux, causing the trough between the peaks to deepen as $N_{\mathrm{HI}}$ increases. This trend is in agreement with the well-known correlation between peak separation and $N_{\mathrm{HI}}$, given the anticorrelation between peak separation and $F_{\mathrm{trough}}/ \bar{F}_{\mathrm{peak}}$ in Figure \ref{fig:sep_trough}.
Therefore, a smaller peak separation and a higher trough flux strongly suggest the presence of low-$N_{\mathrm{HI}}$ channels, which we will explore further in \S \ref{sec:LyC}. 

\subsection{Spatially resolved modeling and variations across halos} 
To characterise the gas properties across the \lya\ halos, we perform spatially resolved spectral modeling using \texttt{zELDA}. We fit spectra from the central `core' region (within a $0.2''$ radius aperture) and the `outer halo' region, selecting either the outermost annular bin or the preceding bin based on spectral quality. The modeling results for four bright halos are presented in Figure~\ref{fig:resolved_examples}. \texttt{zELDA} successfully reproduces the observed double-peaked profiles in both the core and outer halo regions. The presence of a double-peaked profile in the cores indicates that \lya\ photons experience multiple scatterings within the neutral gas shell. This contrasts with the halo studied by \cite{Weldon24}, where the central region appears single-peaked. They attribute this to backscattering, implying strong outflows that allow photons to escape without undergoing additional scatterings. In our case, the persistent double-peaked profile suggests that \lya\ photons are more likely to undergo multiple scatterings before escaping.

For all ten halos, best-fit shell velocities and $N_{\mathrm{HI}}$ values for the core and outer halo regions are shown in Figure~\ref{fig:core_halo}, revealing a clear decline in both parameters from the core to the outskirts. These findings align with previous studies \citep[e.g.,][]{Erb23, Weldon24}, supporting the notion that neutral gas velocity and $N_{\mathrm{HI}}$ regulate \lya\ spectral variations across halos. The \lya\ blue-to-total flux ratio is influenced by the line-of-sight velocity, which is largely a geometric effect arising from the decreasing projected component of the radial outflow velocity with increasing radius \citep{Li22_blob}. Although the decrease in shell expansion velocity in the outermost regions of the \lya\ halo can be attributed to the gravitational deceleration of galactic winds at large radii \cite{Weldon24}, consistent with several observational and simulation studies \citep{Barai13, Thompson16, Chen20}, the broader theoretical picture is more nuanced. Recent high-resolution simulations \citep[e.g.][]{Schneider20} demonstrate that momentum transfer from the hot to the cool phase can efficiently accelerate cool gas out to kiloparsec scales.
Additionally, the enhanced blue peak observed in the outer regions may also result from the preferential scattering of blueshifted \lya\ photons to larger radii \citep{Zheng10}. Peak separation trends are primarily driven by line-of-sight variations in $N_{\mathrm{HI}}$, with lower column densities at larger radii allowing photons to escape closer to the line center without suffering much scattering, reducing overall peak separation. Additionally, the observed trend of increasing residual flux in the absorption troughs with radius is mainly governed by the amount of residual neutral hydrogen in the ISM. A high $N_{\mathrm{HI}}$ indicates a larger amount of \hi\ along the line of sight, which tend to absorb more \lya\ photons near the line center, leading to a deeper trough between the peaks.

\begin{figure}[ht!]
    \centering
    \includegraphics[width=0.95\linewidth]{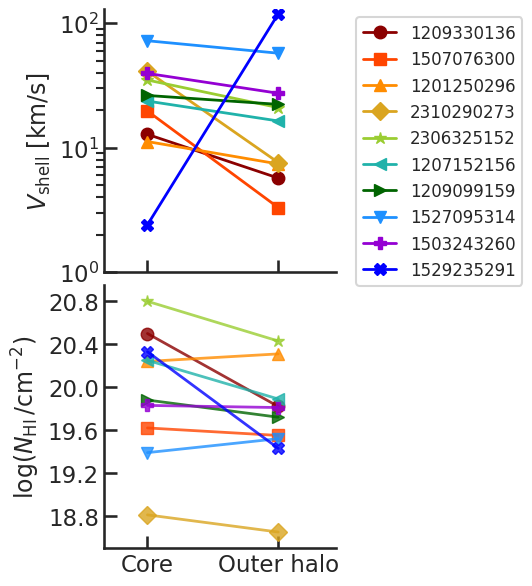}
    \caption{Results from the spatially resolved modeling showing the shell velocity ($V_{\mathrm{shell}}$) and \hi\ column density for two distinct regions: the core and the outer halo. MAGPI IDs are indicated in the legend, following the same order as in Figure~\ref{fig:resolved}.
    }
    \label{fig:core_halo}
\end{figure}

Notably, MAGPI1529235291 shows an increase in shell outflow velocity from the core to outer halo (panel d of Figure~\ref{fig:resolved_examples} and Figure~ \ref{fig:core_halo}), alongside a decreasing flux ratio with radius (Figure~\ref{fig:resolved}). Although its global and central \lya\ spectra are red-dominated, the radial trend resembles that of a blue-dominated (accretion-dominated) system, suggesting active gas accretion in the central region and outflows in the outskirts. This LAE, with an integrated blue-to-total flux ratio of $0.42$, exhibits a higher \lya\ luminosity ($6.92 \times 10^{42}\, \mathrm{erg}\, \mathrm{s}^{-1}$) compared to others within the same ratio bin (see Figure~\ref{fig:lum}), highlighting its distinct behaviour relative to typical outflow-dominated systems. In the future, spatially resolved studies of sources with comparable blue-to-total flux ratios could constrain the turnover point, beyond which the opposite trend is expected.

\subsection{Connection to global LyC escape}\label{sec:LyC}
Previous studies suggest that significant \lya\ flux near systemic velocity may be a potential indicator of LyC escape \citep[e.g.,][]{RT17, Naidu22}. Given that many LAEs in our sample display positive residual flux at the absorption trough or close to the systemic velocity, we revisit the spatially integrated \lya\ profiles to assess their potential as LyC leaker candidates. In the local universe, double peak separation correlates with LyC escape, with narrower peak separations generally associated with higher escape fractions \citep{Verhamme17, Izotov18}. Peak separation is a well-established parameter in the context of \lya\ radiative transfer, and is directly linked to $N_{\mathrm{HI}}$ \citep{Verhamme15, Verhamme17, Li22}. However, peak separation alone does not uniquely reveal \lya\ escape routes \citep{KG21}, as narrow separations can result from multiple small openings in a turbulent medium or a single large, wind-driven cavity \citep{Hu23}. Instead, it can distinguish between different LyC leakage paths. Low LyC leakage, where only a few pathways allow LyC to escape, scattered \lya\ photons traverse optically thick channels with high $N_{\mathrm{HI}}$, resulting in a broad peak separation. In contrast, high leakage corresponds to narrow separations in density-bounded channels. However, at high redshifts ($z > 2$), no clear correlation is observed between peak separation and LyC escape fraction \citep{Kerutt24}.

We present the red peak asymmetry parameter, $A_{f}$, originally introduced by \cite{Rhoads2003}. This parameter has recently been used to quantify the multiphase nature of turbulent \hii\ regions and their connection to LyC escape \citep{KG21, Hu23}.

\begin{figure}[ht!]
    \centering
    \includegraphics[width=0.99\linewidth]{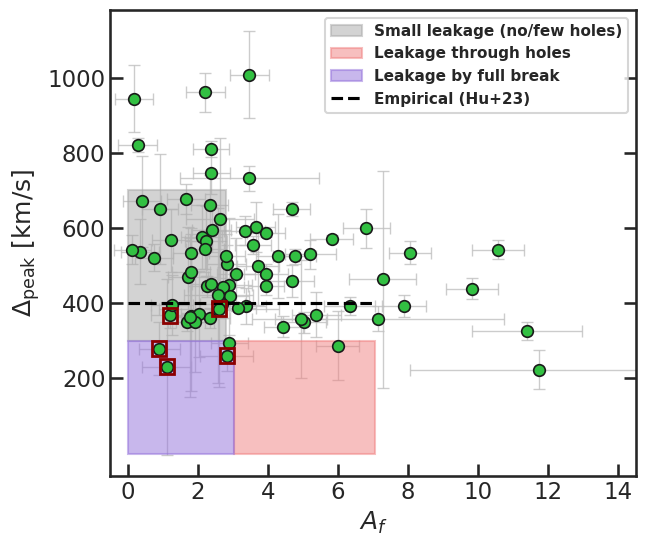}
    \caption{\lya\ double peak separation versus \lya\ red peak asymmetry. The gray, blue, and pink shaded regions correspond to distinct LyC leakage regimes: minimal leakage (gray), escape through low $N_{\mathrm{HI}}$ channels (pink), and escape through large, density-bounded holes (blue). The black dashed line corresponds to the empirical boundary set by \cite{Hu23}. The brown squares mark 5 LAEs in our sample with $F_{\mathrm{trough}}/ \bar{F}_{\mathrm{peak}} \gtrsim 0.2$, indicating they may be potential LyC leakers.
    }
    \label{fig:A_asym}
\end{figure}

Figure~\ref{fig:A_asym} illustrates the relationship between \lya\ peak separation and red peak asymmetry, revealing an anticorrelation: profiles with strong asymmetry exhibit small peak separations, while a wider range of separations is observed for profiles with low asymmetry. When \lya\ photons have direct escape routes through ionised, low-density channels, they experience fewer scatterings, leading to more symmetric profiles. Similarly, if the ISM is largely neutral but has a few clear, direct paths, the result is also relatively symmetric. However, in media with both density-bounded (low $N_{\mathrm{HI}}$) and optically thick neutral regions, some photons escape directly, while others scatter extensively within the dense, neutral gas, resulting in asymmetric profiles. In such a scenario, the scattered photons take longer paths, potentially emerging at different velocities than those escaping directly.


Following \citet{Hu23}, we classify the diagram (Figure~\ref{fig:A_asym}) into three regions: (gray) minimal LyC leakage corresponding to systems with no or very few escape paths; (pink) significant leakage through ionisation-bounded channels with low column density (\fesc $> 10\%$); and (blue) substantial leakage through large holes or density-bounded channels (\fesc $> 10\%$). Based on this classification, we identify five LAEs as potential LyC leakers: four in the blue region ($A_{f} < 3$), suggesting escape through density-bounded channels, and one in the pink region ($A_{f} > 3$), indicating ionisation-bounded leakage in a multiphase medium. Using the empirical relation between peak separation and \fesc\ from \citet{Izotov18}, these five LAEs have \fesc\ $\sim 10 - 27\%$. Many LAEs reside in the gray region, suggesting lower LyC escape fractions compared to those five LAEs in the blue and pink zones. Additionally, as discussed in Section~\ref{sec:4.2.5}, an anticorrelation exists between the peak separation and $F_{\mathrm{trough}}/ \bar{F}_{\mathrm{peak}}$. Three LAEs in the blue zone exhibit higher net \lya\ flux in the trough (i.e., $F_{\mathrm{trough}}/\bar{F}_{\mathrm{peak}} > 0.2$), supporting the conclusion that these galaxies have LyC-thin sightlines \citep{Verhamme15, Gazagnes20}.  Recent studies show that some high-redshift strong leakers instead exhibit large peak separations  \citep[$\Delta_{\mathrm{peak}} > 350\, \mathrm{km} \,\mathrm{s}^{-1}$;][]{Fletcher19, Naidu22, Kerutt24}. \citet{Hu23} has established an empirical boundary between the blue and gray zones, located near a peak separation of 400 km s$^{-1}$. Applying this boundary, along with the trough flux criterion of $F_{\mathrm{trough}}/\bar{F}_{\mathrm{peak}} > 0.2$, we identify an additional two LAEs, resulting in a total of five LAEs as possible LyC leakers.

Cosmological hydrodynamic simulations highlight that the central escape fraction (fraction of the flux near the systemic velocity or absorption trough) is a more robust diagnostic of LyC leakage, whereas peak separation or asymmetry separately are insufficient indicators \citep{Choustikov24}. Given the diagnostic potential of double-peaked \lya\ profiles in constraining CGM kinematics, column density, and LyC escape, expanding double-peaked sample will provide a more statistically significant dataset to further explore the physical mechanisms driving \lya\ and LyC radiative transfer.

\subsection{Caveats in modeling and analysis}\label{sec:caveats}

Finally, we note several caveats, particularly concerning the modeling of \lya\ profiles and the interpretation of the derived results. Our analysis employs a spherically symmetric radiative transfer model applied to annularly averaged spectra, assuming that all \lya\ photons originate at the galaxy center and then scatter through an expanding or infalling shell before escaping into the CGM. This geometry is highly simplified, and the assumptions of central photon production and spherical symmetry are not fully consistent with the setup of annular averaging. In addition, the best-fit models are mainly driven by the high-S/N spectra from the innermost bins, while the outermost bins, though critical for tracing gradients in spectroscopic properties, provide weaker constraints on model parameters. Extracting physical insights from different halo regions will require more realistic anisotropic radiative transfer models with non-uniform \hi\ distributions. Thus, the results presented here should be regarded as average parameters representative of the modeled regions rather than as precise local measurements. 

\section{Summary and conclusions}\label{sec:summary}

We present a comprehensive spectroscopic study of 108 double-peaked LAEs in the first 35 fields of the MAGPI survey, including spatially resolved analysis of ten extended \lya\ halos. To interpret the observed features in the context of gas kinematics in galaxies, we conduct radiative transfer modeling on both global and spatially resolved spectra. Our main findings are summarised as follows:
\begin{itemize}
    \item We apply an automated peak identification technique followed by spatial verification of double- and triple-peaked profiles, identifying 108 double-peaked LAEs. The number of double-peaked LAEs peaks at $z = 3.2$ and drops sharply beyond $z > 4$, likely due to IGM attenuation. The double-peaked fraction at $z < 4$ is $\sim 37$\%, consistent with previous such MUSE results at similar redshifts. Systemic redshifts are determined for 14 LAEs using \ciii\ emission.

    \item We employ a forward modeling approach in which the intrinsic double-asymmetric Gaussian profile is convolved with the instrumental LSF to account for instrumental broadening. This allows us to recover the intrinsic spectra prior to broadening.

    \item We find that 17\% of the double-peaked LAEs are blue-dominated, which may indicate gas accretion, although backscattering remains a possible alternative for sources without systemic redshifts. The blue-to-total flux ratio strongly depends on the \lya\ integrated total S/N. We observe an anti-correlation between \lya\ luminosity and the blue-to-total flux ratio, with fainter LAEs showing a higher fraction of flux in the blue peak. However, the relation flattens at lower luminosities and higher flux ratios.

    \item The spectroscopic properties evolve with redshift: the red peak narrows significantly at $z > 4$, while the blue peak shows a weaker trend. Since the blue peak escape is strongly regulated by the IGM, its persistence implies that the red peak narrowing is instead driven by intrinsic galaxy/CGM evolution (e.g., reduced outflow velocity dispersion, lower \hi\ column densities, or more ionized outflows). We therefore conclude that the \hi\ column density in the ISM and/or CGM is the dominant factor shaping the \lya\ red peak, independent of outflow geometry or kinematics.

    \item Several LAEs in our sample show residual trough flux between the two \lya\ peaks. The normalised trough flux density anticorrelates with the peak separation, both tracing \hi\ column density. The trough flux also correlates with the blue-to-total flux ratio, a trend largely driven by the total line S/N.

    \item We analyse spatially averaged \lya\ profiles across galactic radii in ten LAEs with extended \lya\ halos. Most of them show increasing blue-to-total flux ratios, normalized trough flux densities, and decreasing peak separations with radius. However, these trends are not always spherically symmetric—some systems exhibit significant azimuthal variations, with localized blue flux enhancements and reduced peak separations, complicating interpretations based on radial averages. Spatially-resolved radiative transfer modeling shows a clear line-of-sight effect: a declining shell expansion velocity and \hi\ column density from center to outskirts, explaining the observed trends in blue-to-total flux ratio, peak separation and trough flux densities. One outlier shows increasing shell velocity and decreasing flux ratio in the outer halo. Though red-dominated, its kinematics resemble blue-dominated systems, suggesting a complex structure with central accretion and peripheral outflows.

    \item We observe an inverse correlation between peak separation and red peak asymmetry.  Highly asymmetric red peaks are found to exhibit smaller peak separations. Using peak separation, red peak asymmetry, and normalised trough flux, we identify 5 LAEs in our sample as strong LyC leaker candidates.
\end{itemize}

In general, we find that double-peaked \lya\ profiles, both globally and locally, consistently trace the kinematics and distribution of neutral hydrogen. This demonstrates the power of such profiles in probing the gas dynamics and ionising photon escape in high-redshift galaxies. In the future, a more robust statistical analysis of the double-peak fraction and its redshift evolution will help to further quantify these trends. Additionally, advanced radiative transfer modeling that relaxes the assumption of azimuthal symmetry could provide deeper insight into the complex interplay between double-peaked profiles, gas kinematics, and escape of ionising photons in realistic, anisotropic environments in extended \lya\ halos.


\begin{acknowledgement}

We thank the anonymous referee for their constructive and insightful feedback, which significantly improved this paper.
We thank the ESO staff, and in particular the staff at Paranal Observatory, for carrying out the MAGPI observations. MAGPI targets were selected from GAMA. GAMA is a joint European-Australasian project based around a spectroscopic campaign using the Anglo-Australian Telescope. GAMA was funded by the STFC (UK), the ARC (Australia), the AAO, and the participating institutions. GAMA photometry is based on observations made with ESO Telescopes at the La Silla Paranal Observatory under programme ID 179.A-2004, ID 177.A-3016. The MAGPI team acknowledge support from the Australian Research Council Centre of Excellence for All Sky Astrophysics in 3 Dimensions (ASTRO 3D), through project number CE170100013. K.G. and T.N. acknowledge support from Australian Research Council Laureate Fellowship FL180100060.
SGL acknowledges the financial support from the
MICIU with funding from the European Union NextGenerationEU and Generalitat Valenciana in the call Programa de Planes Complementarios de
I+D+i (PRTR 2022) Project (VAL-JPAS), reference ASFAE/2022/025. This work is part of the research Project PID2023-149420NB-I00 funded by MI-
CIU/AEI/10.13039/501100011033 and by ERDF/EU. This work is also supported by the project of excellence PROMETEO CIPROM/2023/21 of the
Conselleria de Educación, Cultura, Universidades y Empleo (Generalitat Valenciana). CF is the recipient of an Australian Research Council Future Fellowship (project number FT210100168) funded by the Australian Government. CL, JTM and CF are the recipients of funding from the Australian Research Council (ARC) Discovery Project DP210101945. KH acknowledges support by the Royal Society through a Dorothy Hodgkin Fellowship to KA Oman (DHF/R1/231105). SB acknowledges the support from the Physics Foundation through the Messel Research Fellowship. SMS acknowledges funding from the Australian Research Council (DE220100003). LMV acknowledges support by the German Academic Scholarship Foundation (Studienstiftung des deutschen Volkes) and the Marianne-Plehn-Program of the Elite Network of Bavaria.\\

\textbf{Data Availability Statement:} All MUSE data used in this work are publicly accessible through the ESO archive (\href{http://archive.eso.org/cms.html}{http://archive.eso.org/cms.html}). Data products such as fully reduced data cubes and emission line fits will be
made available as part of an upcoming MAGPI team data release (Mendel et al. in prep; Battisti et al. in prep).

\end{acknowledgement}

\printendnotes
\bibliography{pasa}


\appendix
\section{}\label{appendix}
\begin{figure*}
    \centering
    \includegraphics[width=18cm, height=22cm]{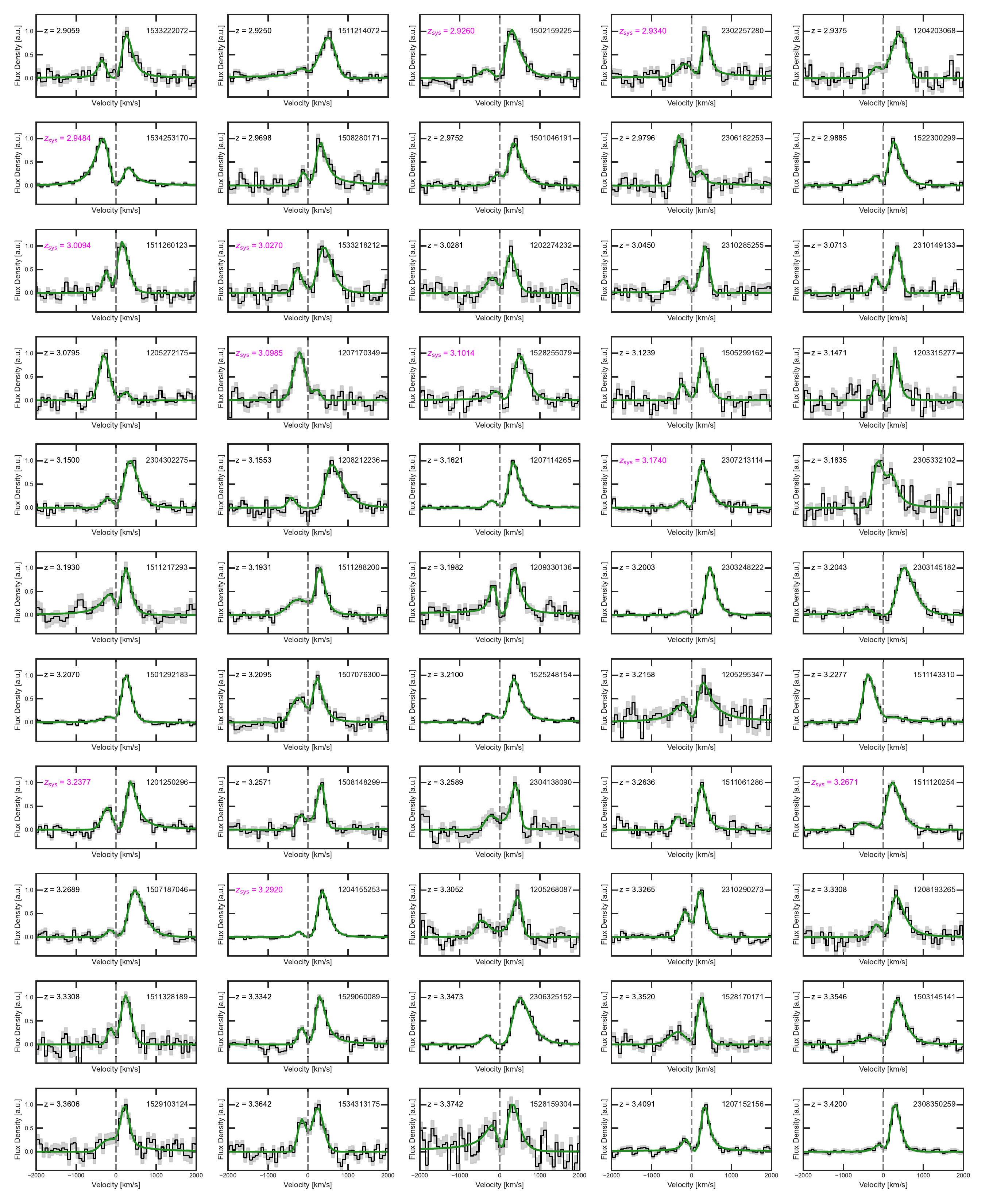}
    \caption{Double-asymmetric Gaussian (DAG) fitting to each line profiles as discussed in \S\ref{sec:forward}. The original spectra are shown in black, while the best-fit convolved models are presented in green. The corresponding MAGPI IDs are shown in the top right corner, and the redshifts are displayed in the top left corner of each frame. For LAEs with a known systemic redshift, the redshifts are highlighted in magenta. The frames are arranged in ascending order of redshift. (Continued)}
    \label{fig:DAG-fit}
\end{figure*}

\begin{figure*}[ht!]
    \centering
    \includegraphics[width=18cm, height=22cm]{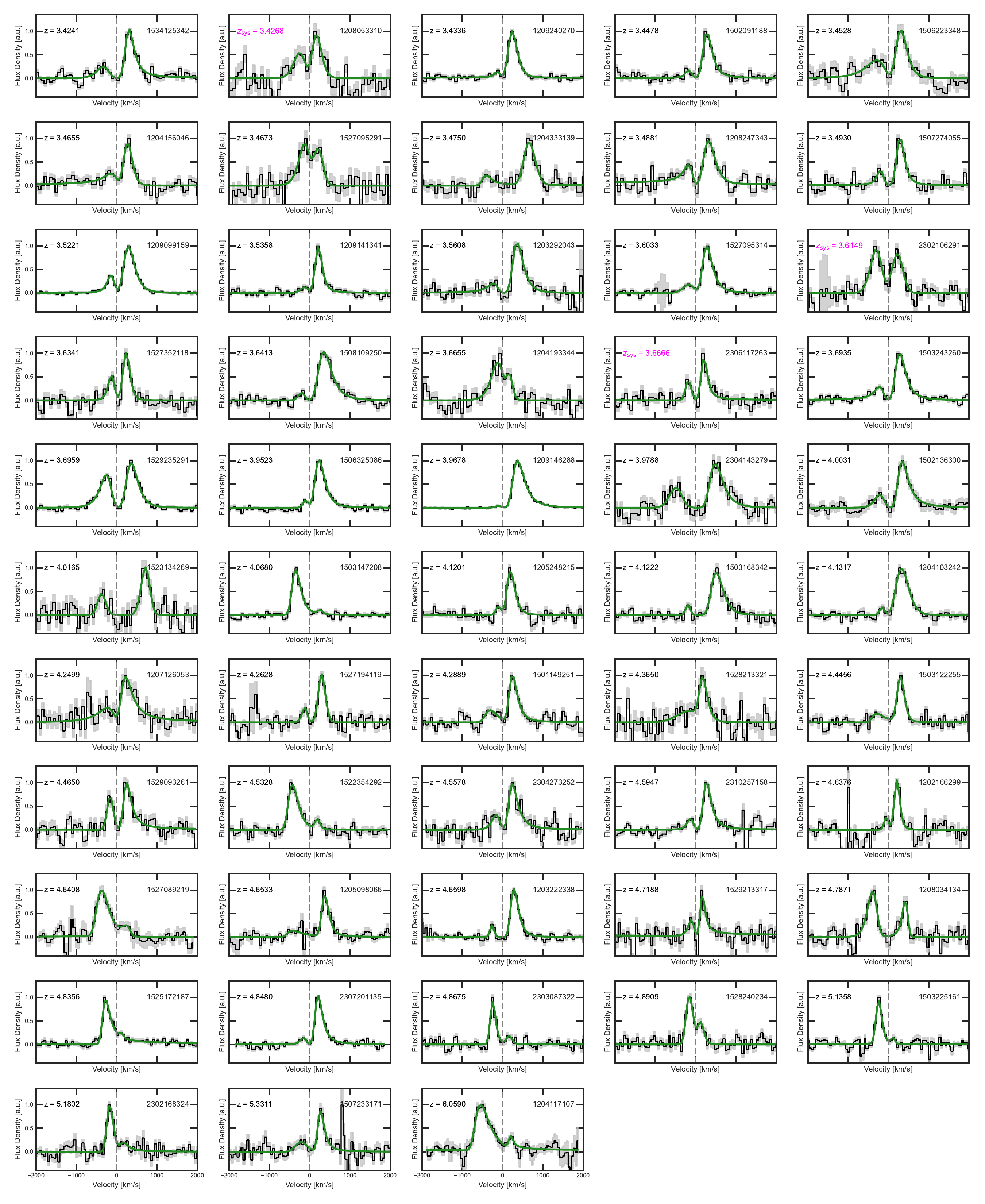}
   \caption{Same as Figure \ref{fig:DAG-fit}.}
   \label{fig:DAG-fit2}
\end{figure*}

\begin{figure*}[h!]
\centering
    {\includegraphics[width=4.2cm,height=3.6cm]{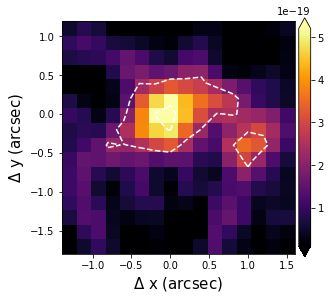}}
    {\includegraphics[width=4.2cm,height=3.6cm]{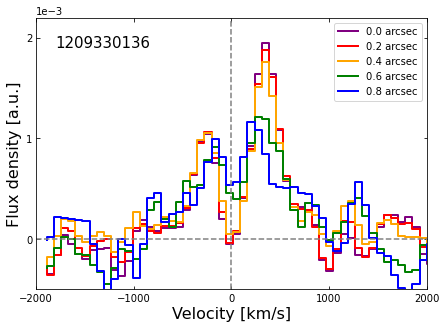}}
    {\includegraphics[width=4.1cm,height=3.6cm]{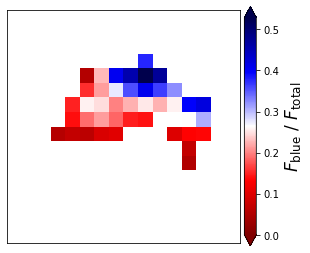}}
    {\includegraphics[width=4.1cm,height=3.6cm]{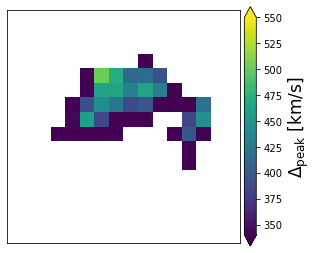}}
    \medskip
    {\includegraphics[width=4.2cm,height=3.6cm]{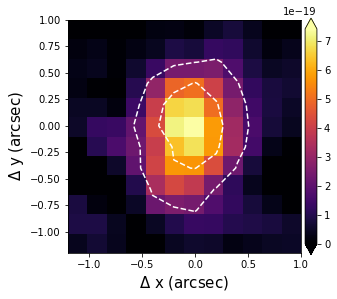}}
    {\includegraphics[width=4.2cm,height=3.6cm]{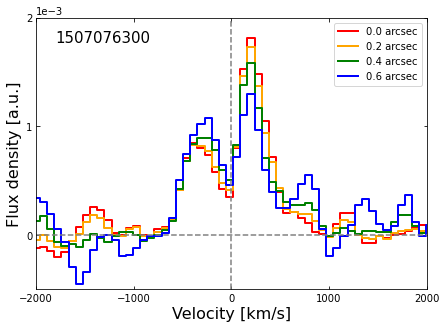}}
    {\includegraphics[width=4.1cm,height=3.6cm]{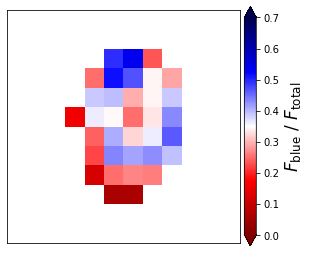}}
    {\includegraphics[width=4.1cm,height=3.6cm]{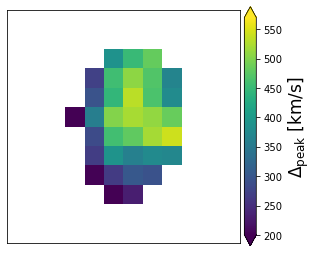}}
    \medskip
    {\includegraphics[width=4.2cm,height=3.6cm]{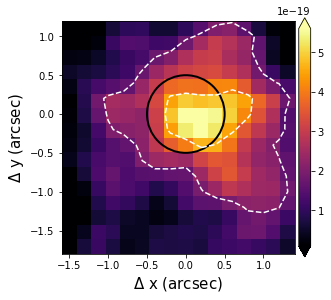}}
    {\includegraphics[width=4.2cm,height=3.6cm]{Figure2/1201250296_radial.png}}
    {\includegraphics[width=4.1cm,height=3.6cm]{Figure2/1201250296_ratio.png}}
    {\includegraphics[width=4.1cm,height=3.6cm]{Figure2/1201250296_sep.png}}
    \medskip
    {\includegraphics[width=4.2cm,height=3.6cm]{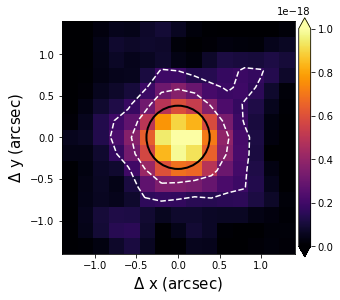}}
    {\includegraphics[width=4.2cm,height=3.6cm]{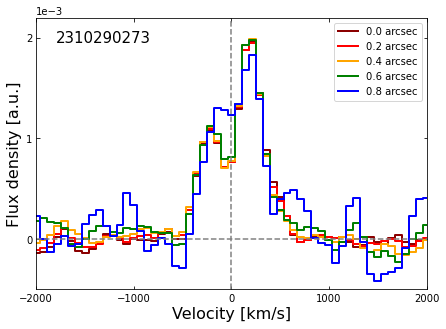}}
    {\includegraphics[width=4.1cm,height=3.6cm]{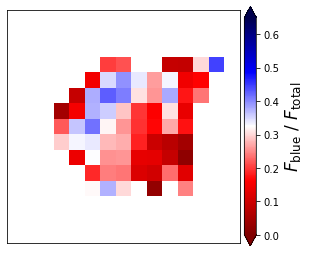}}
    {\includegraphics[width=4.1cm,height=3.6cm]{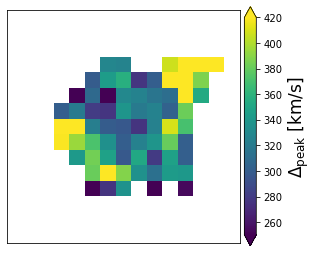}}
    \medskip
    {\includegraphics[width=4.2cm,height=3.6cm]{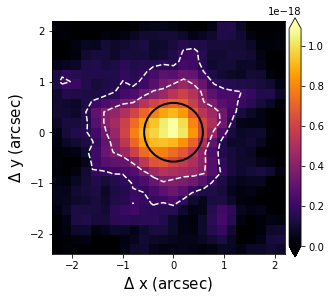}}
    {\includegraphics[width=4.2cm,height=3.6cm]{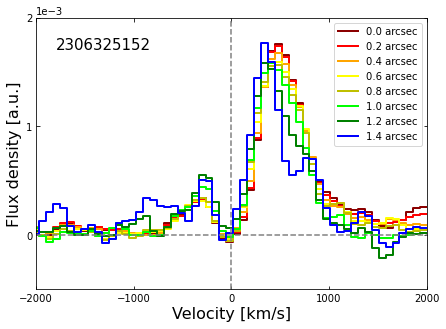}}
    {\includegraphics[width=4.1cm,height=3.6cm]{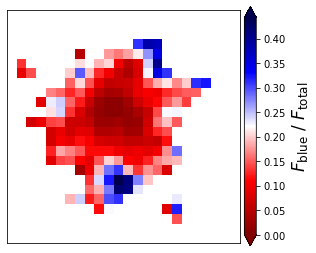}}
    {\includegraphics[width=4.1cm,height=3.6cm]{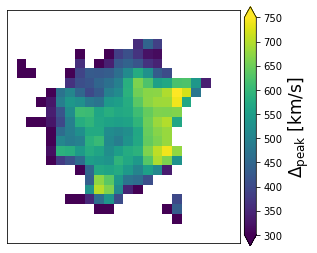}}
    \medskip
    {\includegraphics[width=4.2cm,height=3.6cm]{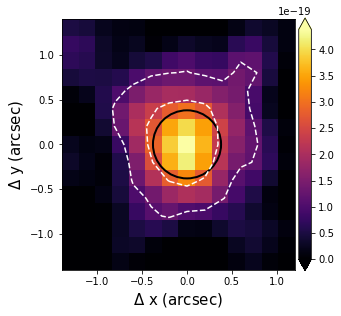}}
    {\includegraphics[width=4.2cm,height=3.6cm]{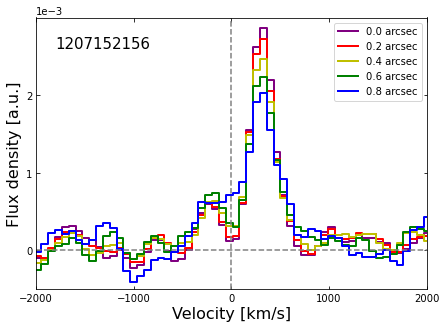}}
    {\includegraphics[width=4.1cm,height=3.6cm]{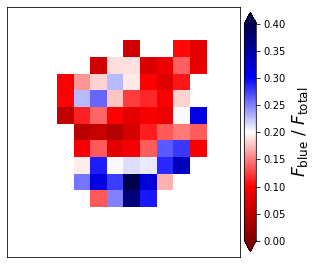}}
    {\includegraphics[width=4.1cm,height=3.6cm]{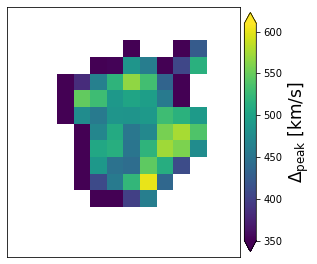}}
    \medskip
\caption{Spatially resolved properties of extended \lya\ halos in our sample. 1st panel: \lya\ pseudo narrow-band image; Coordinates are centred on the brightest pixel. Shown contours represent 2 and 4 $\sigma$ significance levels. The black circles denote the location of stellar continuum detected in the MUSE white light image. 2nd panel: Spectra from annular bins as described in \S \ref{5.2}. Corresponding MAGPI IDs are shown in the top left corner. 3rd panel: blue-to-total flux ratio ($F_{\mathrm{blue}}\, / F_{\mathrm{total}}$) map. 4th panel: Peak separation map. (Continues...)}
\label{SR_properties}
\end{figure*}

\begin{figure*}[ht!]
    \centering 
    {\includegraphics[width=4.2cm,height=3.6cm]{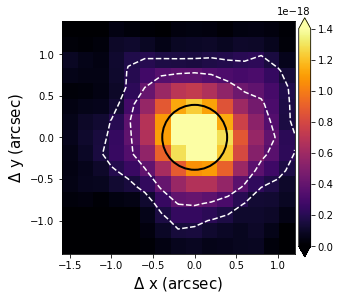}}
    {\includegraphics[width=4.2cm,height=3.6cm]{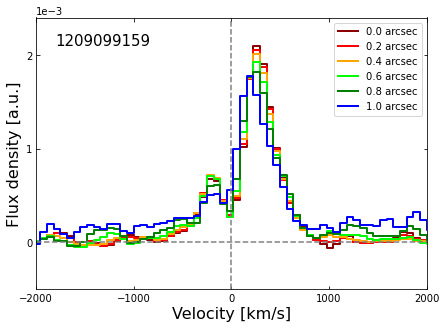}}
    {\includegraphics[width=4.1cm,height=3.6cm]{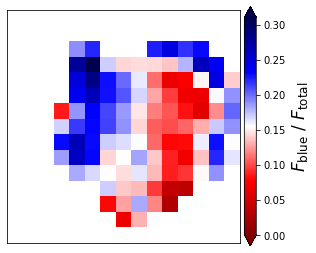}}
    {\includegraphics[width=4.1cm,height=3.6cm]{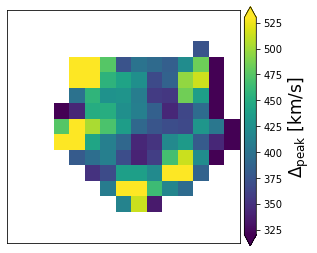}}
    \medskip
    {\includegraphics[width=4.2cm,height=3.6cm]{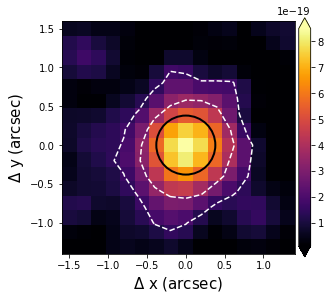}}
    {\includegraphics[width=4.2cm,height=3.6cm]{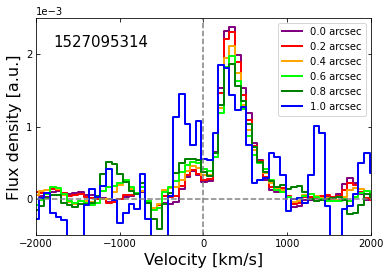}}
    {\includegraphics[width=4.1cm,height=3.6cm]{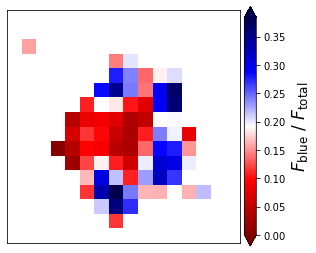}}
    {\includegraphics[width=4.1cm,height=3.6cm]{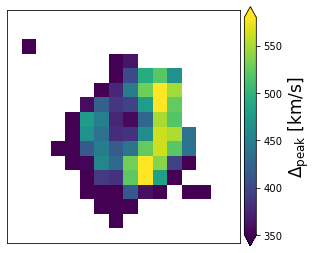}}
    \medskip
    {\includegraphics[width=4.2cm,height=3.6cm]{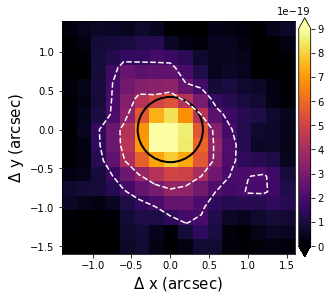}}
    {\includegraphics[width=4.2cm,height=3.6cm]{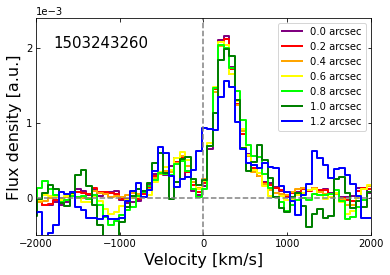}}
    {\includegraphics[width=4.1cm,height=3.6cm]{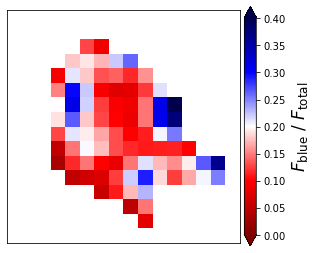}}
    {\includegraphics[width=4.1cm,height=3.6cm]{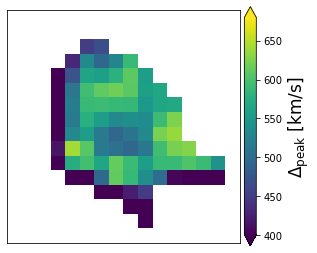}}
    \medskip
    {\includegraphics[width=4.2cm,height=3.6cm]{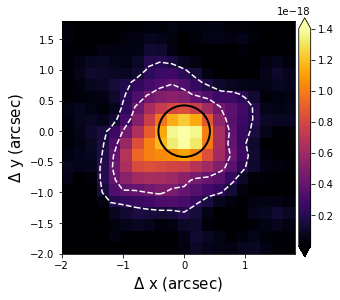}}
    {\includegraphics[width=4.2cm,height=3.6cm]{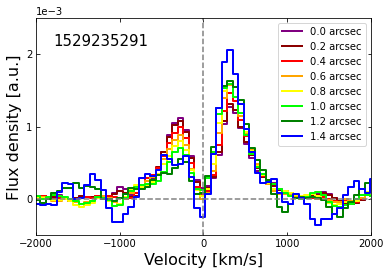}}
    {\includegraphics[width=4.1cm,height=3.6cm]{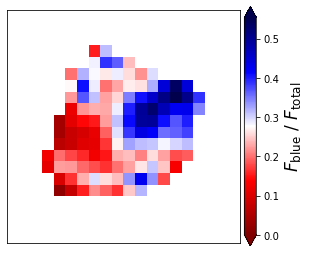}}
    {\includegraphics[width=4.1cm,height=3.6cm]{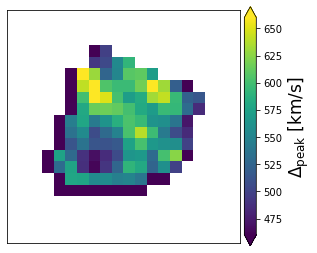}}
\caption{Same as Figure \ref{SR_properties}.}
\end{figure*}


\clearpage
\begin{table*}[hbt!]
\caption{Spectroscopic properties of $110$ MAGPI double-peaked LAEs in the sample, in order of increasing redshift. Columns are as follows: Serial No.; MAGPI ID; $z$: Redshift; $\mathrm{log}\, (L_{\mathrm{Ly}\alpha})$: $\mathrm{Ly}\alpha$ luminosity in $\mathrm{erg}\, \mathrm{s}^{-1}$; $F_{\mathrm{blue}} \,/{F_{\mathrm{total}}}$: Blue-to-total flux ratio; $\Delta_{\mathrm{peak}}$: peak velocity separation, in $\mathrm{km}\,\mathrm{s}^{-1}$ ; $\mathrm{FWHM}_{\,\mathrm{blue}}$: line width (FWHM) of blue $\mathrm{Ly}\alpha$ peak, in $\mathrm{km}\,\mathrm{s}^{-1}$; $\mathrm{FWHM}_{\,\mathrm{red}}$: FWHM of red $\mathrm{Ly}\alpha$ peak, in $\mathrm{km}\,\mathrm{s}^{-1}$; $F_{\mathrm{trough}}/ \bar{F}_{\mathrm{peak}}$: normalised trough flux density; $A_{f}$: \lya\ red peak asymmetry.}
\label{t1}
\begin{tabular}{lccccccccccc}
\toprule
\headrow Sl. No. & MAGPI ID & $z$ & log ($L_{\mathrm{Ly}\alpha}$) & $F_{\mathrm{blue}} \,/{F_{\mathrm{total}}}$  &  $\Delta_{\mathrm{peak}}$  & $\mathrm{FWHM}_{\,\mathrm{blue}}$ & $\mathrm{FWHM}_{\,\mathrm{red}}$ & $F_{\mathrm{trough}}/ \bar{F}_{\mathrm{peak}}$ & $A_{f}$ \\
 & & &  $[\mathrm{erg}\,\mathrm{s}^{-1}]$ &  & $[\mathrm{km}\,\mathrm{s}^{-1}]$ & $[\mathrm{km}\,\mathrm{s}^{-1}]$ & $[\mathrm{km}\,\mathrm{s}^{-1}]$ &  &  \\
\midrule
1 & 1533222072 & $ 2.9059 $ & $ 42.03 \pm 0.05 $ & $ 0.20 \pm 0.07 $ & $ 529.9 \pm 39.4 $ & $ 53.0 \pm 119.5 $ & $ 209.0 \pm 54.7 $ & $ < 0.009 $ & $ 5.21 \pm 0.721 $ \\
2 & 1511214072 & $ 2.9250 $ & $ 42.40 \pm 0.04 $ & $ 0.26 \pm 0.08 $ & $ 651.3 \pm 145.9 $ & $ 303.1 \pm 454.8 $ & $ 424.3 \pm 102.1 $ & $ 0.107 \pm 0.090 $ & $ 0.93 \pm 0.514 $ \\
3 & 1502159225 & $ 2.9260 $ & $ 42.54 \pm 0.03 $ & $ 0.14 \pm 0.05 $ & $ 603.1 \pm 67.5 $ & $ 309.1 \pm 116.4 $ & $ 349.3 \pm 28.9 $ & $ < 0.009 $ & $ 3.65 \pm 0.554 $ \\
4 & 2302257280 & $ 2.9340 $ & $ 41.86 \pm 0.05 $ & $ 0.28 \pm 0.10 $ & $ 498.0 \pm 103.8 $ & $ 380.0 \pm 223.3 $ & $ 114.2 \pm 30.0 $ & $ < 0.009 $ & $ 3.72 \pm 0.590 $ \\
5 & 1204203068 & $ 2.9375 $ & $ 42.12 \pm 0.06 $ & $ 0.14 \pm 0.09 $ & $ 627.4 \pm 253.9 $ & $ 139.8 \pm 308.8 $ & $ 392.1 \pm 119.1 $ & $ 0.194 \pm 0.157 $ & $ 0.73 \pm 0.527 $ \\
6 & 1534253170 & $ 2.9484 $ & $ 42.65 \pm 0.01 $ & $ 0.77 \pm 0.03 $ & $ 571.6 \pm 9.8 $ & $ 337.6 \pm 14.2 $ & $ 250.5 \pm 35.8 $ & $ < 0.009 $ & $ 5.81 \pm 0.608 $ \\
7 & 1508280171 & $ 2.9698 $ & $ 42.03 \pm 0.06 $ & $ 0.13 \pm 0.06 $ & $ 384.6 \pm 153.7 $ & $ 51.1 \pm 307.7 $ & $ 215.5 \pm 66.1 $ & $ < 0.009 $ & $ 5.59 \pm 0.652 $ \\
8 & 1501046191 & $ 2.9752 $ & $ 42.55 \pm 0.04 $ & $ 0.17 \pm 0.06 $ & $ 366.8 \pm 199.4 $ & $ 263.5 \pm 260.3 $ & $ 273.0 \pm 74.4 $ & $ 0.145 \pm 0.089 $ & $ 1.81 \pm 0.532 $ \\
9 & 2306182253 & $ 2.9796 $ & $ 41.94 \pm 0.05 $ & $ 0.83 \pm 0.12 $ & $ 573.9 \pm 179.7 $ & $ 262.2 \pm 64.6 $ & $ 111.4 \pm 269.6 $ & $ 0.165 \pm 0.133 $ & $ 1.31 \pm 0.737 $ \\
10 & 1522300299 & $ 2.9885 $ & $ 42.45 \pm 0.02 $ & $ 0.16 \pm 0.05 $ & $ 393.0 \pm 28.9 $ & $ 168.6 \pm 97.6 $ & $ 257.4 \pm 20.6 $ & $ < 0.009 $ & $ 7.88 \pm 0.633 $ \\
11 & 1511260123 & $ 3.0094 $ & $ 42.26 \pm 0.04 $ & $ 0.23 \pm 0.06 $ & $ 349.2 \pm 27.4 $ & $ 75.1 \pm 59.9 $ & $ 243.3 \pm 30.8 $ & $ < 0.009 $ & $ 5.02 \pm 0.650 $ \\
12 & 1533218212 & $ 3.0270 $ & $ 42.07 \pm 0.05 $ & $ 0.22 \pm 0.09 $ & $ 677.8 \pm 40.7 $ & $ 155.1 \pm 75.2 $ & $ 366.0 \pm 52.2 $ & $ 0.044 \pm 0.114 $ & $ 1.66 \pm 0.545 $ \\
13 & 1202274232 & $ 3.0281 $ & $ 41.90 \pm 0.07 $ & $ 0.30 \pm 0.11 $ & $ 448.8 \pm 177.4 $ & $ 264.4 \pm 202.2 $ & $ 215.1 \pm 70.6 $ & $ 0.059 \pm 0.164 $ & $ 2.87 \pm 0.664 $ \\
14 & 2310285255 & $ 3.0450 $ & $ 42.00 \pm 0.05 $ & $ 0.25 \pm 0.09 $ & $ 536.5 \pm 86.3 $ & $ 109.7 \pm 279.3 $ & $ 131.6 \pm 25.1 $ & $ 0.023 \pm 0.079 $ & $ 0.36 \pm 0.542 $ \\
15 & 2310149133 & $ 3.0713 $ & $ 42.14 \pm 0.04 $ & $ 0.21 \pm 0.06 $ & $ 542.7 \pm 38.3 $ & $ 83.8 \pm 68.3 $ & $ 166.4 \pm 31.8 $ & $ 0.011 \pm 0.067 $ & $ 0.13 \pm 0.525 $ \\
16 & 1205272175 & $ 3.0795 $ & $ 42.18 \pm 0.04 $ & $ 0.88 \pm 0.11 $ & $ 565.0 \pm 251.8 $ & $ 247.6 \pm 111.6 $ & $ 77.6 \pm 348.6 $ & $ 0.038 \pm 0.087 $ & $ 0.84 \pm 0.941 $ \\
17 & 1207170349 & $ 3.0985 $ & $ 42.07 \pm 0.05 $ & $ 0.91 \pm 0.14 $ & $ 483.9 \pm 270.7 $ & $ 283.6 \pm 147.4 $ & $ 95.6 \pm 345.1 $ & $ 0.128 \pm 0.136 $ & $ 0.18 \pm 0.706 $ \\
18 & 1528255079 & $ 3.1014 $ & $ 42.11 \pm 0.04 $ & $ 0.10 \pm 0.06 $ & $ 554.0 \pm 104.8 $ & $ 159.1 \pm 332.0 $ & $ 355.8 \pm 34.2 $ & $ < 0.009 $ & $ 1.29 \pm 0.534 $ \\
19 & 1505299162 & $ 3.1239 $ & $ 41.98 \pm 0.05 $ & $ 0.20 \pm 0.08 $ & $ 529.2 \pm 118.6 $ & $ 116.2 \pm 188.0 $ & $ 193.7 \pm 63.6 $ & $ 0.026 \pm 0.118 $ & $ 1.57 \pm 0.570 $ \\
20 & 1203315277 & $ 3.1471 $ & $ 41.83 \pm 0.09 $ & $ 0.22 \pm 0.12 $ & $ 461.2 \pm 235.1 $ & $ 74.0 \pm 161.0 $ & $ 100.8 \pm 39.0 $ & $ < 0.009 $ & $ 3.92 \pm 1.950 $ \\
21 & 2304302275 & $ 3.1500 $ & $ 42.34 \pm 0.03 $ & $ 0.11 \pm 0.04 $ & $ 534.3 \pm 49.7 $ & $ 189.9 \pm 134.8 $ & $ 359.2 \pm 26.3 $ & $ < 0.009 $ & $ 1.80 \pm 0.524 $ \\
22 & 1208212236 & $ 3.1553 $ & $ 42.32 \pm 0.03 $ & $ 0.10 \pm 0.03 $ & $ 1008.5 \pm 115.2 $ & $ 133.0 \pm 65.0 $ & $ 448.6 \pm 36.5 $ & $ < 0.009 $ & $ 3.47 \pm 0.564 $ \\
23 & 1207114265 & $ 3.1621 $ & $ 42.49 \pm 0.02 $ & $ 0.10 \pm 0.03 $ & $ 534.3 \pm 31.2 $ & $ 199.8 \pm 62.9 $ & $ 239.6 \pm 15.0 $ & $ 0.012 \pm 0.036 $ & $ 8.05 \pm 0.597 $ \\
24 & 2307213114 & $ 3.1740 $ & $ 42.01 \pm 0.03 $ & $ 0.12 \pm 0.04 $ & $ 476.5 \pm 96.0 $ & $ 107.4 \pm 152.1 $ & $ 267.9 \pm 21.0 $ & $ < 0.009 $ & $ 3.09 \pm 0.551 $ \\
25 & 2305332102 & $ 3.1835 $ & $ 42.02 \pm 0.08 $ & $ 0.64 \pm 0.17 $ & $ 330.2 \pm 299.0 $ & $ 296.8 \pm 199.6 $ & $ 207.4 \pm 268.5 $ & $ 0.725 \pm 0.262 $ & $ 2.57 \pm 0.631 $ \\
26 & 1511217293 & $ 3.1930 $ & $ 41.97 \pm 0.06 $ & $ 0.36 \pm 0.12 $ & $ 358.9 \pm 157.2 $ & $ 336.6 \pm 369.1 $ & $ 185.0 \pm 64.7 $ & $ 0.034 \pm 0.131 $ & $ 4.94 \pm 0.740 $ \\
27 & 1511288200 & $ 3.1931 $ & $ 42.42 \pm 0.06 $ & $ 0.48 \pm 0.12 $ & $ 505.2 \pm 49.9 $ & $ 673.8 \pm 107.0 $ & $ 187.4 \pm 33.5 $ & $ 0.261 \pm 0.080 $ & $ 2.82 \pm 0.562 $ \\
28 & 1209330136 & $ 3.1982 $ & $ 42.20 \pm 0.04 $ & $ 0.31 \pm 0.07 $ & $ 477.5 \pm 26.8 $ & $ 73.7 \pm 49.2 $ & $ 238.1 \pm 59.2 $ & $ < 0.009 $ & $ 3.93 \pm 0.625 $ \\
29 & 2303248222 & $ 3.2003 $ & $ 42.50 \pm 0.02 $ & $ 0.04 \pm 0.02 $ & $ 585.5 \pm 305.0 $ & $ 52.5 \pm 351.0 $ & $ 219.9 \pm 17.2 $ & $ < 0.009 $ & $ 5.20 \pm 0.592 $ \\
30 & 2303145182 & $ 3.2043 $ & $ 42.39 \pm 0.02 $ & $ 0.07 \pm 0.03 $ & $ 891.4 \pm 128.7 $ & $ 371.6 \pm 331.9 $ & $ 451.5 \pm 26.5 $ & $ < 0.009 $ & $ 2.10 \pm 0.528 $ \\
31 & 1501292183 & $ 3.2070 $ & $ 42.62 \pm 0.02 $ & $ 0.12 \pm 0.03 $ & $ 370.5 \pm 114.5 $ & $ 284.1 \pm 228.7 $ & $ 224.1 \pm 17.3 $ & $ 0.030 \pm 0.037 $ & $ 2.04 \pm 0.518 $ \\
32 & 1507076300 & $ 3.2095 $ & $ 42.32 \pm 0.06 $ & $ 0.46 \pm 0.11 $ & $ 458.4 \pm 42.7 $ & $ 403.7 \pm 86.0 $ & $ 211.1 \pm 50.4 $ & $ 0.214 \pm 0.103 $ & $ 4.69 \pm 0.620 $ \\
33 & 1525248154 & $ 3.2100 $ & $ 42.73 \pm 0.01 $ & $ 0.12 \pm 0.02 $ & $ 650.5 \pm 19.0 $ & $ 307.7 \pm 94.4 $ & $ 286.9 \pm 14.5 $ & $ 0.087 \pm 0.034 $ & $ 4.68 \pm 0.525 $ \\
34 & 1205295347 & $ 3.2158 $ & $ 41.98 \pm 0.07 $ & $ 0.26 \pm 0.11 $ & $ 455.9 \pm 182.8 $ & $ 399.2 \pm 369.7 $ & $ 349.6 \pm 101.6 $ & $ 0.019 \pm 0.196 $ & $ 5.58 \pm 0.694 $ \\
35 & 1511143310 & $ 3.2277 $ & $ 42.47 \pm 0.03 $ & $ 0.88 \pm 0.07 $ & $ 674.3 \pm 531.4 $ & $ 289.4 \pm 41.8 $ & $ 524.8 \pm 471.3 $ & $ 0.034 \pm 0.066 $ & $ 4.09 \pm 0.671 $ \\
36 & 1201250296 & $ 3.2377 $ & $ 42.40 \pm 0.02 $ & $ 0.28 \pm 0.03 $ & $ 542.7 \pm 24.4 $ & $ 229.7 \pm 36.4 $ & $ 244.7 \pm 27.3 $ & $ < 0.009 $ & $ 10.56 \pm 0.730 $ \\
37 & 1508148299 & $ 3.2571 $ & $ 42.09 \pm 0.07 $ & $ 0.27 \pm 0.09 $ & $ 566.9 \pm 83.9 $ & $ 205.2 \pm 242.5 $ & $ 154.7 \pm 40.3 $ & $ 0.182 \pm 0.084 $ & $ 1.23 \pm 0.550 $ \\
38 & 2304138090 & $ 3.2589 $ & $ 41.98 \pm 0.09 $ & $ 0.46 \pm 0.18 $ & $ 671.3 \pm 121.1 $ & $ 371.6 \pm 249.6 $ & $ 151.0 \pm 47.6 $ & $ 0.209 \pm 0.155 $ & $ 0.40 \pm 0.550 $ \\
39 & 1511061286 & $ 3.2636 $ & $ 42.27 \pm 0.03 $ & $ 0.23 \pm 0.05 $ & $ 592.4 \pm 38.5 $ & $ 229.8 \pm 101.6 $ & $ 192.6 \pm 34.6 $ & $ 0.053 \pm 0.078 $ & $ 3.34 \pm 0.587 $ \\
40 & 1511120254 & $ 3.2671 $ & $ 42.60 \pm 0.02 $ & $ 0.13 \pm 0.03 $ & $ 746.3 \pm 55.4 $ & $ 327.0 \pm 81.2 $ & $ 370.2 \pm 16.8 $ & $ 0.037 \pm 0.059 $ & $ 2.38 \pm 0.526 $ \\
41 & 1507187046 & $ 3.2689 $ & $ 42.36 \pm 0.02 $ & $ 0.06 \pm 0.03 $ & $ 620.5 \pm 251.3 $ & $ 159.9 \pm 298.8 $ & $ 407.8 \pm 36.5 $ & $ 0.017 \pm 0.101 $ & $ 2.65 \pm 0.538 $ \\
42 & 1204155253 & $ 3.2920 $ & $ 42.68 \pm 0.01 $ & $ 0.06 \pm 0.01 $ & $ 554.1 \pm 24.0 $ & $ 129.5 \pm 50.7 $ & $ 261.5 \pm 9.7 $ & $ < 0.009 $ & $ 3.57 \pm 0.522 $ \\
43 & 1205268087 & $ 3.3052 $ & $ 42.02 \pm 0.08 $ & $ 0.34 \pm 0.12 $ & $ 944.4 \pm 89.8 $ & $ 343.4 \pm 198.1 $ & $ 213.5 \pm 68.4 $ & $ 0.024 \pm 0.146 $ & $ 0.18 \pm 0.531 $ \\
44 & 2310290273 & $ 3.3265 $ & $ 42.48 \pm 0.03 $ & $ 0.35 \pm 0.05 $ & $ 349.4 \pm 23.2 $ & $ 227.2 \pm 45.9 $ & $ 185.2 \pm 27.1 $ & $ 0.079 \pm 0.047 $ & $ 1.69 \pm 0.528 $ \\
45 & 1208193265 & $ 3.3308 $ & $ 42.14 \pm 0.05 $ & $ 0.15 \pm 0.07 $ & $ 461.6 \pm 277.0 $ & $ 189.6 \pm 241.3 $ & $ 334.6 \pm 88.3 $ & $ 0.016 \pm 0.147 $ & $ 5.10 \pm 0.660 $ \\
\bottomrule
\end{tabular}
\end{table*}

\begin{table*}[hbt!]
\caption{Same as in Table \ref{t1}}
\label{t2}
\begin{tabular}{lccccccccc}
\toprule
\headrow Sl. No. & MAGPI ID & $z$ & log ($L_{\mathrm{Ly}\alpha}$) & $F_{\mathrm{blue}} \,/{F_{\mathrm{total}}}$  &  $\Delta_{\mathrm{peak}}$  & $\mathrm{FWHM}_{\,\mathrm{blue}}$ & $\mathrm{FWHM}_{\,\mathrm{red}}$ & $F_{\mathrm{trough}}/ \bar{F}_{\mathrm{peak}}$ & $A_{f}$ \\
 & & &  $[\mathrm{erg}\,\mathrm{s}^{-1}]$ &  & $[\mathrm{km}\,\mathrm{s}^{-1}]$ & $[\mathrm{km}\,\mathrm{s}^{-1}]$ & $[\mathrm{km}\,\mathrm{s}^{-1}]$ &  &  \\
\midrule
46 & 1511328189 & $ 3.3308 $ & $ 42.00 \pm 0.05 $ & $ 0.16 \pm 0.08 $ & $ 376.3 \pm 439.4 $ & $ 65.5 \pm 316.6 $ & $ 246.2 \pm 116.6 $ & $ < 0.009 $ & $ 2.77 \pm 0.624 $ \\
47 & 1529060089 & $ 3.3342 $ & $ 42.37 \pm 0.02 $ & $ 0.18 \pm 0.03 $ & $ 438.6 \pm 27.7 $ & $ 144.1 \pm 54.5 $ & $ 233.9 \pm 30.0 $ & $ 0.013 \pm 0.060 $ & $ 9.82 \pm 0.737 $ \\
48 & 2306325152 & $ 3.3473 $ & $ 42.79 \pm 0.01 $ & $ 0.10 \pm 0.01 $ & $ 810.4 \pm 21.1 $ & $ 214.5 \pm 31.9 $ & $ 463.8 \pm 10.9 $ & $ < 0.009 $ & $ 2.38 \pm 0.513 $ \\
49 & 1528170171 & $ 3.3520 $ & $ 42.09 \pm 0.05 $ & $ 0.30 \pm 0.09 $ & $ 575.6 \pm 85.1 $ & $ 431.3 \pm 146.8 $ & $ 197.3 \pm 31.1 $ & $ 0.036 \pm 0.113 $ & $ 2.11 \pm 0.587 $ \\
50 & 1503145141 & $ 3.3546 $ & $ 42.50 \pm 0.03 $ & $ 0.15 \pm 0.04 $ & $ 660.1 \pm 82.1 $ & $ 577.9 \pm 186.1 $ & $ 340.2 \pm 24.6 $ & $ 0.073 \pm 0.079 $ & $ 2.34 \pm 0.528 $ \\
51 & 1529103124 & $ 3.3606 $ & $ 41.95 \pm 0.11 $ & $ 0.24 \pm 0.16 $ & $ 342.6 \pm 485.9 $ & $ 821.3 \pm 430.4 $ & $ 185.7 \pm 98.9 $ & $ 0.383 \pm 0.214 $ & $ 2.78 \pm 0.595 $ \\
52 & 1534313175 & $ 3.3642 $ & $ 42.17 \pm 0.05 $ & $ 0.35 \pm 0.09 $ & $ 368.0 \pm 35.0 $ & $ 194.6 \pm 73.1 $ & $ 272.5 \pm 61.5 $ & $ 0.375 \pm 0.107 $ & $ 1.20 \pm 0.543 $ \\
53 & 1528159304 & $ 3.3742 $ & $ 41.93 \pm 0.06 $ & $ 0.35 \pm 0.12 $ & $ 469.8 \pm 132.0 $ & $ 356.1 \pm 299.6 $ & $ 299.3 \pm 69.8 $ & $ < 0.009 $ & $ 1.72 \pm 0.571 $ \\
54 & 1207152156 & $ 3.4091 $ & $ 42.19 \pm 0.03 $ & $ 0.18 \pm 0.04 $ & $ 392.4 \pm 47.6 $ & $ 106.3 \pm 100.0 $ & $ 160.0 \pm 31.0 $ & $ < 0.009 $ & $ 3.37 \pm 0.560 $ \\
55 & 2308350259 & $ 3.4200 $ & $ 42.71 \pm 0.02 $ & $ 0.09 \pm 0.03 $ & $ 350.4 \pm 132.6 $ & $ 205.7 \pm 337.4 $ & $ 214.1 \pm 23.1 $ & $ 0.019 \pm 0.058 $ & $ 1.92 \pm 0.532 $ \\
56 & 1534125342 & $ 3.4241 $ & $ 42.39 \pm 0.03 $ & $ 0.19 \pm 0.05 $ & $ 599.6 \pm 51.8 $ & $ 316.4 \pm 100.6 $ & $ 186.3 \pm 28.1 $ & $ < 0.009 $ & $ 6.81 \pm 0.667 $ \\
57 & 1208053310 & $ 3.4268 $ & $ 42.13 \pm 0.09 $ & $ 0.43 \pm 0.17 $ & $ 396.0 \pm 79.7 $ & $ 282.1 \pm 166.5 $ & $ 217.3 \pm 90.7 $ & $ 0.127 \pm 0.194 $ & $ 1.26 \pm 0.609 $ \\
58 & 1209240270 & $ 3.4336 $ & $ 42.43 \pm 0.02 $ & $ 0.08 \pm 0.03 $ & $ 287.3 \pm 92.4 $ & $ 57.6 \pm 220.8 $ & $ 200.5 \pm 19.2 $ & $ < 0.009 $ & $ 5.99 \pm 0.607 $ \\
59 & 1502091188 & $ 3.4478 $ & $ 42.36 \pm 0.03 $ & $ 0.08 \pm 0.04 $ & $ 385.5 \pm 338.8 $ & $ 59.4 \pm 384.3 $ & $ 199.4 \pm 40.8 $ & $ < 0.009 $ & $ 2.48 \pm 0.557 $ \\
60 & 1506223348 & $ 3.4528 $ & $ 42.27 \pm 0.06 $ & $ 0.37 \pm 0.11 $ & $ 565.8 \pm 92.8 $ & $ 516.0 \pm 229.4 $ & $ 271.0 \pm 52.0 $ & $ 0.021 \pm 0.172 $ & $ 2.24 \pm 0.599 $ \\
61 & 1204156046 & $ 3.4655 $ & $ 42.15 \pm 0.05 $ & $ 0.22 \pm 0.08 $ & $ 362.1 \pm 210.6 $ & $ 405.7 \pm 355.2 $ & $ 181.7 \pm 47.2 $ & $ < 0.009 $ & $ 1.77 \pm 0.569 $ \\
62 & 1527095291 & $ 3.4673 $ & $ 42.07 \pm 0.10 $ & $ 0.48 \pm 0.19 $ & $ 279.2 \pm 71.3 $ & $ 191.8 \pm 156.1 $ & $ 250.9 \pm 149.2 $ & $ 0.305 \pm 0.151 $ & $ 0.90 \pm 0.585 $ \\
63 & 1204333139 & $ 3.4750 $ & $ 42.09 \pm 0.05 $ & $ 0.13 \pm 0.09 $ & $ 1088.3 \pm 428.0 $ & $ 213.0 \pm 338.0 $ & $ 287.4 \pm 55.2 $ & $ 0.038 \pm 0.160 $ & $ 1.26 \pm 0.546 $ \\
64 & 1208247343 & $ 3.4881 $ & $ 42.32 \pm 0.04 $ & $ 0.29 \pm 0.08 $ & $ 445.2 \pm 27.6 $ & $ 175.8 \pm 86.5 $ & $ 287.3 \pm 39.0 $ & $ < 0.009 $ & $ 2.26 \pm 0.548 $ \\
65 & 1507274055 & $ 3.4930 $ & $ 42.16 \pm 0.04 $ & $ 0.21 \pm 0.06 $ & $ 442.4 \pm 53.2 $ & $ 56.5 \pm 153.2 $ & $ 159.2 \pm 39.0 $ & $ < 0.009 $ & $ 2.71 \pm 0.601 $ \\
66 & 1209099159 & $ 3.5221 $ & $ 42.82 \pm 0.01 $ & $ 0.20 \pm 0.01 $ & $ 388.0 \pm 7.9 $ & $ 125.0 \pm 18.7 $ & $ 268.9 \pm 9.2 $ & $ < 0.009 $ & $ 3.15 \pm 0.520 $ \\
67 & 1209141341 & $ 3.5358 $ & $ 42.46 \pm 0.02 $ & $ 0.11 \pm 0.04 $ & $ 297.8 \pm 250.7 $ & $ 132.7 \pm 329.7 $ & $ 114.5 \pm 23.6 $ & $ < 0.009 $ & $ 10.55 \pm 0.880 $ \\
68 & 1203292043 & $ 3.5608 $ & $ 42.54 \pm 0.04 $ & $ 0.15 \pm 0.07 $ & $ 542.2 \pm 125.0 $ & $ 258.4 \pm 258.6 $ & $ 254.5 \pm 27.6 $ & $ < 0.009 $ & $ 2.46 \pm 0.607 $ \\
69 & 1527095314 & $ 3.6033 $ & $ 42.58 \pm 0.03 $ & $ 0.15 \pm 0.04 $ & $ 420.2 \pm 39.6 $ & $ 274.5 \pm 95.7 $ & $ 238.2 \pm 17.9 $ & $ 0.029 \pm 0.053 $ & $ 2.93 \pm 0.533 $ \\
70 & 2302106291 & $ 3.6149 $ & $ 42.25 \pm 0.04 $ & $ 0.53 \pm 0.09 $ & $ 521.0 \pm 36.6 $ & $ 289.2 \pm 51.4 $ & $ 279.2 \pm 50.1 $ & $ 0.149 \pm 0.131 $ & $ 0.75 \pm 0.546 $ \\
71 & 1527352118 & $ 3.6341 $ & $ 42.47 \pm 0.05 $ & $ 0.35 \pm 0.07 $ & $ 325.8 \pm 24.2 $ & $ 106.8 \pm 50.5 $ & $ 124.3 \pm 24.2 $ & $ < 0.009 $ & $ 11.40 \pm 1.573 $ \\
72 & 1508109250 & $ 3.6413 $ & $ 42.46 \pm 0.02 $ & $ 0.06 \pm 0.02 $ & $ 526.7 \pm 58.3 $ & $ 100.8 \pm 91.2 $ & $ 355.7 \pm 20.6 $ & $ < 0.009 $ & $ 2.80 \pm 0.537 $ \\
73 & 1204193344 & $ 3.6655 $ & $ 42.23 \pm 0.11 $ & $ 0.67 \pm 0.26 $ & $ 230.3 \pm 234.8 $ & $ 309.0 \pm 185.2 $ & $ 91.5 \pm 110.6 $ & $ 0.223 \pm 0.118 $ & $ 1.11 \pm 0.706 $ \\
74 & 2306117263 & $ 3.6666 $ & $ 41.98 \pm 0.09 $ & $ 0.25 \pm 0.09 $ & $ 359.3 \pm 32.5 $ & $ 71.7 \pm 50.9 $ & $ 101.3 \pm 38.8 $ & $ < 0.009 $ & $ 7.14 \pm 3.586 $ \\
75 & 1503243260 & $ 3.6935 $ & $ 42.62 \pm 0.02 $ & $ 0.18 \pm 0.04 $ & $ 450.0 \pm 20.2 $ & $ 213.8 \pm 78.8 $ & $ 248.9 \pm 18.0 $ & $ < 0.009 $ & $ 2.37 \pm 0.525 $ \\
76 & 1529235291 & $ 3.6959 $ & $ 42.84 \pm 0.01 $ & $ 0.42 \pm 0.02 $ & $ 586.1 \pm 9.0 $ & $ 280.1 \pm 17.6 $ & $ 290.5 \pm 13.3 $ & $ < 0.009 $ & $ 3.95 \pm 0.539 $ \\
77 & 1506325086 & $ 3.9523 $ & $ 42.81 \pm 0.02 $ & $ 0.10 \pm 0.02 $ & $ 293.4 \pm 20.8 $ & $ 51.1 \pm 27.6 $ & $ 238.5 \pm 12.5 $ & $ < 0.009 $ & $ 2.90 \pm 0.529 $ \\
78 & 1209146288 & $ 3.9678 $ & $ 43.30 \pm 0.00 $ & $ 0.02 \pm 0.01 $ & $ 446.0 \pm 25.5 $ & $ 67.6 \pm 49.8 $ & $ 308.8 \pm 4.1 $ & $ < 0.009 $ & $ 3.94 \pm 0.511 $ \\
79 & 2304143279 & $ 3.9788 $ & $ 42.39 \pm 0.04 $ & $ 0.32 \pm 0.07 $ & $ 961.1 \pm 52.3 $ & $ 332.5 \pm 72.4 $ & $ 355.9 \pm 45.4 $ & $ < 0.009 $ & $ 2.22 \pm 0.564 $ \\
80 & 1502136300 & $ 4.0031 $ & $ 42.33 \pm 0.03 $ & $ 0.13 \pm 0.05 $ & $ 525.4 \pm 112.1 $ & $ 233.6 \pm 210.5 $ & $ 267.2 \pm 45.4 $ & $ < 0.009 $ & $ 4.27 \pm 0.621 $ \\
81 & 1523134269 & $ 4.0165 $ & $ 42.09 \pm 0.11 $ & $ 0.35 \pm 0.25 $ & $ 1098.8 \pm 54.2 $ & $ 193.0 \pm 115.7 $ & $ 216.4 \pm 45.7 $ & $ < 0.009 $ & $ 0.45 \pm 0.552 $ \\
82 & 1503147208 & $ 4.0680 $ & $ 42.59 \pm 0.03 $ & $ 0.88 \pm 0.08 $ & $ 619.3 \pm 483.3 $ & $ 205.8 \pm 55.7 $ & $ 47.0 \pm 215.4 $ & $ 0.060 \pm 0.051 $ & $ 3.23 \pm 1.007 $ \\
83 & 1205248215 & $ 4.1201 $ & $ 42.28 \pm 0.04 $ & $ 0.10 \pm 0.05 $ & $ 277.2 \pm 170.7 $ & $ 49.6 \pm 153.7 $ & $ 182.4 \pm 28.2 $ & $ < 0.009 $ & $ 1.48 \pm 0.552 $ \\
84 & 1503168342 & $ 4.1222 $ & $ 42.41 \pm 0.03 $ & $ 0.08 \pm 0.03 $ & $ 690.0 \pm 35.0 $ & $ 34.0 \pm 41.3 $ & $ 314.9 \pm 28.3 $ & $ < 0.009 $ & $ 2.09 \pm 0.546 $ \\
85 & 1204103242 & $ 4.1317 $ & $ 42.52 \pm 0.02 $ & $ 0.06 \pm 0.02 $ & $ 482.3 \pm 117.2 $ & $ 47.8 \pm 116.9 $ & $ 343.7 \pm 23.7 $ & $ < 0.009 $ & $ 1.80 \pm 0.534 $ \\
86 & 1207126053 & $ 4.2499 $ & $ 41.99 \pm 0.06 $ & $ 0.20 \pm 0.09 $ & $ 417.2 \pm 312.6 $ & $ 453.5 \pm 376.8 $ & $ 306.7 \pm 126.3 $ & $ 0.067 \pm 0.236 $ & $ 6.21 \pm 0.718 $ \\
87 & 1527194119 & $ 4.2628 $ & $ 42.24 \pm 0.04 $ & $ 0.19 \pm 0.07 $ & $ 398.6 \pm 39.0 $ & $ 89.8 \pm 113.6 $ & $ 161.2 \pm 24.4 $ & $ < 0.009 $ & $ 2.69 \pm 0.640 $ \\
88 & 1501149251 & $ 4.2889 $ & $ 42.36 \pm 0.04 $ & $ 0.24 \pm 0.07 $ & $ 543.0 \pm 50.8 $ & $ 314.4 \pm 82.6 $ & $ 187.0 \pm 23.8 $ & $ 0.054 \pm 0.093 $ & $ 2.20 \pm 0.554 $ \\
89 & 1528213321 & $ 4.3650 $ & $ 42.17 \pm 0.09 $ & $ 0.32 \pm 0.19 $ & $ 306.1 \pm 382.0 $ & $ 721.6 \pm 274.7 $ & $ 181.8 \pm 189.1 $ & $ 0.365 \pm 0.254 $ & $ 1.14 \pm 0.581 $ \\
90 & 1503122255 & $ 4.4456 $ & $ 42.40 \pm 0.04 $ & $ 0.21 \pm 0.07 $ & $ 595.5 \pm 59.6 $ & $ 307.6 \pm 191.5 $ & $ 172.5 \pm 22.3 $ & $ 0.077 \pm 0.086 $ & $ 2.39 \pm 0.565 $ \\
\bottomrule
\end{tabular}
\end{table*}

\begin{table*}[ht!]
\caption{Same as in Table \ref{t1}}
\begin{tabular}{lccccccccccc}
\toprule
\headrow Sl. No. & MAGPI ID & $z$ & log ($L_{\mathrm{Ly}\alpha}$) & $F_{\mathrm{blue}} \,/{F_{\mathrm{total}}}$  &  $\Delta_{\mathrm{peak}}$  & $\mathrm{FWHM}_{\,\mathrm{blue}}$ & $\mathrm{FWHM}_{\,\mathrm{red}}$ & $F_{\mathrm{trough}}/ \bar{F}_{\mathrm{peak}}$ & $A_{f}$ \\
 & & &  $[\mathrm{erg}\,\mathrm{s}^{-1}]$ &  & $[\mathrm{km}\,\mathrm{s}^{-1}]$ & $[\mathrm{km}\,\mathrm{s}^{-1}]$ & $[\mathrm{km}\,\mathrm{s}^{-1}]$ &  &  \\
\midrule
91 & 1529093261 & $ 4.4650 $ & $ 42.26 \pm 0.04 $ & $ 0.33 \pm 0.07 $ & $ 393.0 \pm 24.7 $ & $ 137.0 \pm 37.9 $ & $ 183.4 \pm 41.9 $ & $ < 0.009 $ & $ 6.33 \pm 0.734 $ \\
92 & 1522354292 & $ 4.5328 $ & $ 42.79 \pm 0.03 $ & $ 0.88 \pm 0.08 $ & $ 624.4 \pm 214.1 $ & $ 258.1 \pm 28.4 $ & $ 81.0 \pm 155.3 $ & $ 0.064 \pm 0.100 $ & $ 2.64 \pm 0.972 $ \\
93 & 2304273252 & $ 4.5578 $ & $ 42.26 \pm 0.04 $ & $ 0.20 \pm 0.06 $ & $ 369.6 \pm 59.5 $ & $ 182.2 \pm 158.9 $ & $ 229.6 \pm 40.2 $ & $ < 0.009 $ & $ 5.38 \pm 0.691 $ \\
94 & 2310257158 & $ 4.5947 $ & $ 42.50 \pm 0.02 $ & $ 0.13 \pm 0.03 $ & $ 360.6 \pm 25.5 $ & $ 130.2 \pm 76.2 $ & $ 200.3 \pm 20.0 $ & $ < 0.009 $ & $ 2.35 \pm 0.555 $ \\
95 & 1202166299 & $ 4.6376 $ & $ 42.11 \pm 0.04 $ & $ 0.16 \pm 0.06 $ & $ 241.6 \pm 226.0 $ & $ 47.7 \pm 197.8 $ & $ 103.0 \pm 24.3 $ & $ < 0.009 $ & $ 5.90 \pm 0.887 $ \\
96 & 1527089219 & $ 4.6408 $ & $ 42.46 \pm 0.05 $ & $ 0.75 \pm 0.11 $ & $ 650.6 \pm 725.6 $ & $ 331.7 \pm 45.6 $ & $ 227.4 \pm 359.4 $ & $ 0.329 \pm 0.153 $ & $ 0.30 \pm 0.634 $ \\
97 & 1205098066 & $ 4.6533 $ & $ 42.62 \pm 0.04 $ & $ 0.22 \pm 0.08 $ & $ 786.9 \pm 107.6 $ & $ 411.7 \pm 234.0 $ & $ 164.2 \pm 28.5 $ & $ 0.046 \pm 0.120 $ & $ 1.82 \pm 0.553 $ \\
98 & 1203222338 & $ 4.6598 $ & $ 42.76 \pm 0.02 $ & $ 0.10 \pm 0.03 $ & $ 525.0 \pm 19.5 $ & $ 37.5 \pm 19.1 $ & $ 187.8 \pm 13.9 $ & $ < 0.009 $ & $ 4.76 \pm 0.609 $ \\
99 & 1529213317 & $ 4.7188 $ & $ 42.17 \pm 0.04 $ & $ 0.26 \pm 0.08 $ & $ 223.5 \pm 52.7 $ & $ 111.0 \pm 154.5 $ & $ 72.4 \pm 31.4 $ & $ < 0.009 $ & $ 11.72 \pm 3.677 $ \\
100 & 1208034134 & $ 4.7871 $ & $ 42.78 \pm 0.03 $ & $ 0.67 \pm 0.07 $ & $ 820.2 \pm 18.1 $ & $ 251.4 \pm 27.2 $ & $ 110.4 \pm 35.4 $ & $ 0.015 \pm 0.083 $ & $ 0.28 \pm 0.562 $ \\
101 & 1525172187 & $ 4.8356 $ & $ 42.56 \pm 0.07 $ & $ 0.62 \pm 0.14 $ & $ 384.8 \pm 208.5 $ & $ 191.6 \pm 35.6 $ & $ 63.7 \pm 116.8 $ & $ 0.256 \pm 0.087 $ & $ 2.60 \pm 1.705 $ \\
102 & 2307201135 & $ 4.8480 $ & $ 42.42 \pm 0.02 $ & $ 0.09 \pm 0.02 $ & $ 337.3 \pm 27.9 $ & $ 137.7 \pm 69.0 $ & $ 179.0 \pm 12.1 $ & $ < 0.009 $ & $ 4.44 \pm 0.559 $ \\
103 & 2303087322 & $ 4.8675 $ & $ 42.30 \pm 0.04 $ & $ 0.82 \pm 0.10 $ & $ 420.9 \pm 233.7 $ & $ 99.4 \pm 20.6 $ & $ 126.2 \pm 140.7 $ & $ 0.046 \pm 0.075 $ & $ 2.59 \pm 0.981 $ \\
104 & 1528240234 & $ 4.8909 $ & $ 42.30 \pm 0.06 $ & $ 0.70 \pm 0.15 $ & $ 260.6 \pm 41.8 $ & $ 174.0 \pm 31.3 $ & $ 127.4 \pm 116.7 $ & $ 0.273 \pm 0.107 $ & $ 2.82 \pm 0.757 $ \\
105 & 1503225161 & $ 5.1358 $ & $ 42.50 \pm 0.04 $ & $ 0.90 \pm 0.10 $ & $ 364.1 \pm 369.8 $ & $ 133.5 \pm 23.0 $ & $ 122.3 \pm 95.5 $ & $ 0.093 \pm 0.077 $ & $ 1.09 \pm 1.433 $ \\
106 & 2302168324 & $ 5.1802 $ & $ 42.36 \pm 0.07 $ & $ 0.78 \pm 0.17 $ & $ 339.5 \pm 523.0 $ & $ 156.3 \pm 22.6 $ & $ 165.4 \pm 151.5 $ & $ 0.126 \pm 0.089 $ & $ 8.18 \pm 1.293 $ \\
107 & 1507233171 & $ 5.3311 $ & $ 42.40 \pm 0.05 $ & $ 0.25 \pm 0.09 $ & $ 463.4 \pm 288.4 $ & $ 325.7 \pm 293.4 $ & $ 152.5 \pm 41.3 $ & $ 0.026 \pm 0.137 $ & $ 7.27 \pm 0.955 $ \\
108 & 1204117107 & $ 6.0590 $ & $ 42.88 \pm 0.05 $ & $ 0.84 \pm 0.13 $ & $ 732.5 \pm 32.7 $ & $ 418.7 \pm 30.7 $ & $ 61.6 \pm 69.4 $ & $ 0.168 \pm 0.130 $ & $ 3.47 \pm 1.976 $ \\
\bottomrule
\end{tabular}
\end{table*}

\end{document}